\documentclass[aps,prd,showpacs,floatfix,preprintnumbers,nofootinbib,amssymb,amsmath]{revtex4}
\usepackage{graphicx,bm,color}
\usepackage{subfigure}
\def \l {\left}
\def \r {\right}

\def \bit{\begin{itemize}}
\def \eit{\end{itemize}}
\def \ben{\begin{enumerate}}
\def \een{\end{enumerate}}
\def \bsubeq{\begin{subequations}}
\def \esubeq{\end{subequations}}
\def \beq{\begin{equation}}
\def \eeq{\end{equation}}
\def \beqa{\begin{eqnarray}}
\def \eeqa{\end{eqnarray}}
\def \L{{\mathbf L}}
\newcommand{\cC}{{\cal C}}
\newcommand{\cD}{{\cal D}}

\newcommand{\cP}{{\cal P}}
\newcommand{\cU}{{\cal U}}
\newcommand{\cV}{{\cal V}}

\def \etal{{\sl et al.~\/}}
\def \ibid{{\sl ibid~\/}}

\def \eup{{\sl Eur.\ Phys. ~\/}}
\def \ijmp{{\sl Int.\ J.\ Mod.\ Phys. ~\/}}
\def \jhep{{\sl J.\ H.\ E.\ P.~\/}}

\def \np{{\sl Nucl.\ Phys.~\/}}

\def \pl{{\sl Phys.\ Lett.~\/}}
\def \pr{{\sl Phys.\ Rev.~\/}}
\def \prp{{\sl Phys.\ Rept.~\/}}

\def \zphys{{\sl Z.\ Phys.~\/}}
\def \ie{{\it i.e.\/}}

\def \etal{{\sl et al. \/}}

\begin{document}
 
\title{PNJL model with a Vandermonde term}
\author{Sanjay K.\ \surname{Ghosh}}
\email{sanjay@bosemain.boseinst.ac.in}
\author{Tamal K.\ \surname{Mukherjee}}
\email{tamal@bosemain.boseinst.ac.in}
\affiliation{Department of Physics, Bose Institute, 
          93/1, A.P.C. Road, Kolkata 700 009, India.}
\author{Munshi G. \surname{Mustafa}}
\email{munshigolam.mustafa@saha.ac.in}
\author{Rajarshi \surname{Ray}}
\email{rajarshi.ray@saha.ac.in}
\affiliation{Theory Division, Saha Institute of Nuclear Physics, 
          1/AF, Bidhannagar, Kolkata 700 064, India.}

\begin{abstract}

We extend the Polyakov-Nambu-Jona-Lasinio (PNJL) model  for two
degenerate flavours by including the effect of the SU(3) measure with
a Vandermonde (VdM) term. This ensures that the Polyakov loop always
remains in the domain [0,1]. The pressure, energy density, specific heat,
speed of sound and conformal measure show small or negligible effects
from this term. However various quark number and isospin susceptibilities 
are all found to approach their respective ideal gas limits around 2 $T_c$.
We compare our methods with other similar approaches in PNJL model and also 
present a quantitative comparison with Lattice QCD data.

\end{abstract}
\preprint{SINP/TNP/2007-31}
\maketitle

\section{Introduction} \label{sc.intro}

Recently, there is a lot of interest in the studies of thermodynamics
of strongly interacting matter using the Polyakov loop enhanced 
Nambu-Jona-Lasinio (PNJL) model \cite{pnjl0,pnjl1,pnjl2}. This model 
couples the chiral and deconfinement order parameters through a
simple-minded coupling of the NJL model \cite{njl1} with the Polyakov 
loop model \cite{polyd1}. The two major thrusts in recent times have 
been to estimate various thermodynamic observables using this model 
(see e.g. \cite{pnjl3,mesoni,pnjl4, isospini,susci}), and to make 
systematic improvements of the model \cite{ratti4,polrgi,ringi}. 
Another set of important result has come from similar studies in
chiral quark models that go beyond the mean field treatment 
\cite{enrique}.

In this note we deal with the improvement of the Polyakov loop model
and describe some of its consequences, remaining within the domain of
mean field analysis. The Polyakov loop model used 
in many of the recent literature is the one given in Ref.\cite{pnjl2}. 
The Polyakov loop $\Phi$ has been treated here as a Z(3) spin field 
\cite{polyl}. 
Using this model we estimated \cite{pnjl3} a very sensitive observable 
- the quark number susceptibility (QNS) and also the higher order 
coefficients in the Taylor expansion of pressure in quark number 
chemical potential $\mu_0$. Comparison with the data from Lattice QCD 
(LQCD) \cite{sixx} showed that the QNS in the PNJL model and LQCD 
agree quite well both qualitatively and quantitatively. The fourth 
order coefficient $c_4$ showed qualitative agreement but had a 
quantitative difference at high temperatures. 
Some of us further extended the PNJL model to include isospin chemical
potential $\mu_I$ \cite{pnjl4}. The isospin number susceptibility 
(INS) and its derivative with respect to $\mu_0$ and $\mu_I$
were obtained. In this case the fourth order derivative $c_4^I$ was
quite consistent with lattice data, but the INS was not. A possible 
reason for such departures is that the mean-field treatment of the 
PNJL model is insufficient. But then it should have affected the
coefficients systematically \ie, all the fourth order coefficients
should deviate further from LQCD data than the second order coefficients.

There are however other {\it simpler} reasons that should be 
considered first. The PNJL model is only a model which 
can mimic some of the characteristics of a fundamental theory like QCD 
and its discretized version LQCD. Moreover, the parameters like the 
couplings and masses are quite different in the PNJL model and the 
LQCD simulations. Thus some quantitative difference is naturally 
expected. Apart from these we made an important observation in
\cite{pnjl4} that $\Phi$ has a big role to play in the behaviour of
these coefficients. We pointed out how the quantitative differences
could be caused by the behaviour of $\Phi$ as a function 
of temperature and chemical potentials. The most important 
{\it physical problem} in the simple-minded PNJL model is the
following. $\Phi$ being the normalized trace of the Wilson line 
$\L$, which is an SU(3) matrix, should lie in the range 
$0 \le \Phi \le 1$. But it was found to be greater than 1 at 
temperatures above 2$T_c$ (see Fig.2 in Ref. \cite{pnjl4}). 
The natural way to cure this problem is to consider a 
proper Jacobian of transformation from the matrix valued field $L$ to 
the complex valued field $\Phi$ which will then constrain the value
of $\Phi$ to $\Phi < 1$. This is quite a well known construction in SU(N) matrix 
model (see e.g.\cite{dumitru,steinacker,akemann}), in certain variations
of Polyakov loop model (\cite{meisinger2,ratti4}), as well as in 
QCD motivated phenomenological models (see \cite{mustafa} and references
therein).
Also this is ubiquitous in various strong coupling 
effective theories of Lattice QCD (see e.g. \cite{strc1}).

Here we introduce the Vandermonde term in the Polyakov loop model
in a conceptually different way than that in the earlier models. In 
the next section we discuss our approach. In section III we show 
the changes in measurements of the susceptibilities and various other 
quantities due to the VdM term. The final section contains our 
conclusions.

\section{Formalism} \label{sc.formal}

At a temperature $T$, the SU(3) Wilson line is given by 
$ \L ({\bf x}) = \cP {\rm exp} (ig \int_0^{1/T} A_0^a ({\bf x}) \lambda_a d\tau)$, 
where $g$ is the gauge coupling, $A_0^a$ ($a$ = 1,2,...8) are the 
time-like components of the gluon field, $\lambda_a$ are the Gell-Mann 
matrices and $\tau$ the imaginary time in the Euclidian field theory.
The Polyakov loop is defined as $\Phi = {\rm tr} \L /3$ and its 
conjugate is $\bar{\Phi} = {\rm tr} \L^{\dagger} /3$.
Since $\L$ is itself a SU(3) matrix so $\Phi,\bar{\Phi} \le 1$. The gluon 
thermodynamics can be described as an effective theory of the Polyakov 
loops \cite{polyd1}. On the other hand quark thermodynamics can be 
effectively described in terms of NJL model \cite{njl1}, and the two
are coupled to obtain the PNJL model (e.g., \cite{pnjl2}).
The thermodynamic potential in this model can be obtained in terms of the
sigma and pion condensates and the thermal average of the Polyakov loop.

However the version of the PNJL model \cite{pnjl2} leads to $\Phi > 1$
for $T > 2T_c$. To rectify this anomaly, the authors of Ref.\ \cite{pnjl2}
have recently proposed 
a complete modification of the Polyakov loop model \cite{ratti4}, 
motivated from the strong coupling results used by Fukushima \cite{pnjl1}
Our aim in this work is also similar, but the approach is somewhat 
different. We retain the Polyakov loop potential of \cite{pnjl2,pnjl3,pnjl4} 
but treat it as a matrix model. Also the way we define pressure is quite
different as discussed below.

We first outline our scheme using an arbitrary matrix model for the 
Wilson line $\L$, which for simplicity is assumed to be a potential
$\cV[\L] \equiv \cV[\Phi,\bar{\Phi}]$. In the following equation,
we express the partition function for this theory first as a path
integral over $\L$ and then over the fields $\Phi$ and $\bar{\Phi}$.

\bsubeq
\beqa
Z 
= \int \cD\L \, {\rm e}^{-\,\frac{1}{T}\cV[\Phi,\bar{\Phi}]}
&=& 
\int \prod_{\bf x} d\L({\bf x})\,{\rm e}^{-\,\frac{1}{T}\cV[\Phi,\bar{\Phi}]}
\label{eq.inttrace}\\
&=&
\int \prod_{\bf x} J[\Phi({\bf x}),\bar{\Phi}({\bf x})] \, 
d\Phi({\bf x}) \, d\bar{\Phi}({\bf x}) 
                      \, {\rm e}^{-\,\frac{1}{T}\cV[\Phi,\bar{\Phi}]}
\label{eq.intmes} 
\eeqa
\esubeq

\noindent
where, $\cD \L$ is the SU(3) Haar measure, $J[\Phi,\bar{\Phi}]$ is 
the Jacobian of transformation (also called Vandermonde determinant, 
see e.g. Ref. \cite{trinhammer}) from $\L$ to ($\Phi,\bar{\Phi}$), 
and is given as $J[\Phi,\bar{\Phi}] \equiv (27/24\pi^2)
(1 - 6\,\bar{\Phi} \Phi + 4\,(\bar{\Phi}^3 + \Phi^3) 
- 3\,(\bar{\Phi} \Phi)^2)$. Our interest then would be to obtain
the pressure which is given by,

\beqa
P  =  T \frac{\partial\, \ln Z}{\partial v} 
   = - \left\langle \frac{\partial\, \cV }{\partial v} \right\rangle
\simeq - \frac{1}{v} \langle \cV \rangle 
\label{eq.prest}
\eeqa

\noindent
where, $v$ denotes the physical volume of the system and $\langle \rangle$
denotes thermal averaging. The last approximation holds in the infinite
volume limit. 

The role of the Jacobian is to be understood as follows. First, 
it is a factor reweighting the field configurations and hence 
significantly affects all thermal averages. However the Jacobian
is not explicitly space-time dependent, there is no extra term to be 
averaged in Eqn. \ref{eq.prest} as one might expect when redefining
the path integration from $\L$ to $\Phi$ (Eqn. \ref{eq.inttrace}
to Eqn. \ref{eq.intmes}). 
A typical example of such a dependence
would be if we were considering say a Fourier transform of the
fields. In case of a free field this kind of dependence
of the Jacobian on the volume and temperature is very important in
obtaining the correct partition function.

Thus, in our mean field treatment we have to carefully incorporate
the effect of the Jacobian and this is the main aim of this paper.
The effect of the Jacobian is reflected in the mean fields 
$\left<\Phi\right>$ and $\left<\bar{\Phi}\right>$, and we express
the pressure as,
                
\beqa           
P = - \frac{1}{v}\cV(\left<\Phi\right>,\left<\bar{\Phi}\right>).
\eeqa 

\noindent
To relate to pure glue theory,  we now replace the potential 
density $\cV/v$ by a Landau-Ginzburg type functional $\cU$, 
given by\cite{pnjl2},

\beqa
\frac{\mathcal{U}\left(\Phi,\bar{\Phi},T\right)}{ T^4} =
-\frac{b_2\left(T\right)}{ 2 }\bar{\Phi} \Phi-
\frac{b_3}{6}\left(\Phi^3+
{\bar{\Phi}}^3\right)+ \frac{b_4}{4}\left(\bar{\Phi} \Phi\right)^2 ~~~,
\label{eq.uu}
\eeqa

\noindent
with

\beqa
b_2\left(T\right)=a_0+a_1\left(\frac{T_0}{T}\right)
+a_2\left(\frac{T_0}{T} \right)^2+a_3\left(\frac{T_0}{T}\right)^3~~~.
\label{eq.bb}
\eeqa

\begin{figure}[!tbh]
\subfigure{
   {\includegraphics [scale=0.6] {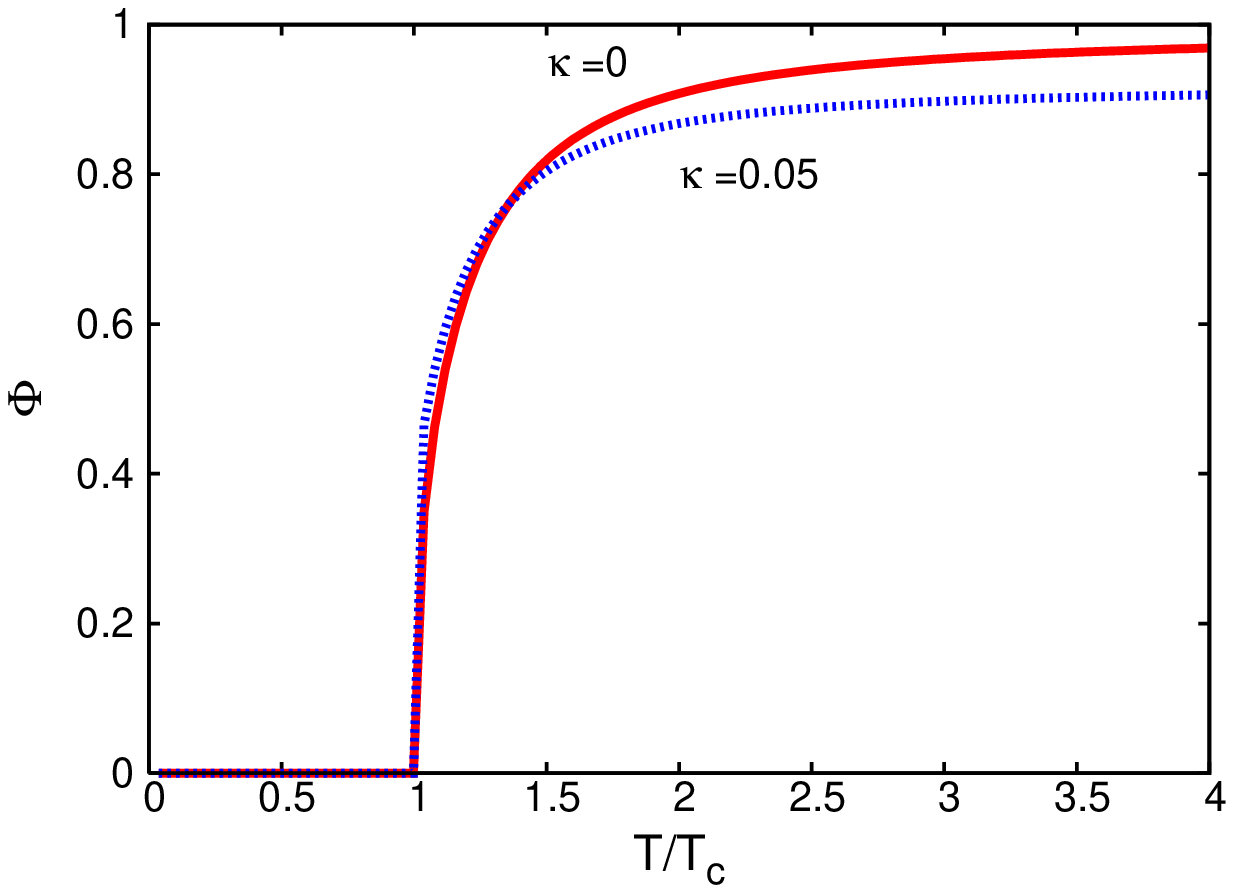}}
}
\hskip 0.15 in
\subfigure{
   {\includegraphics[scale=0.6]{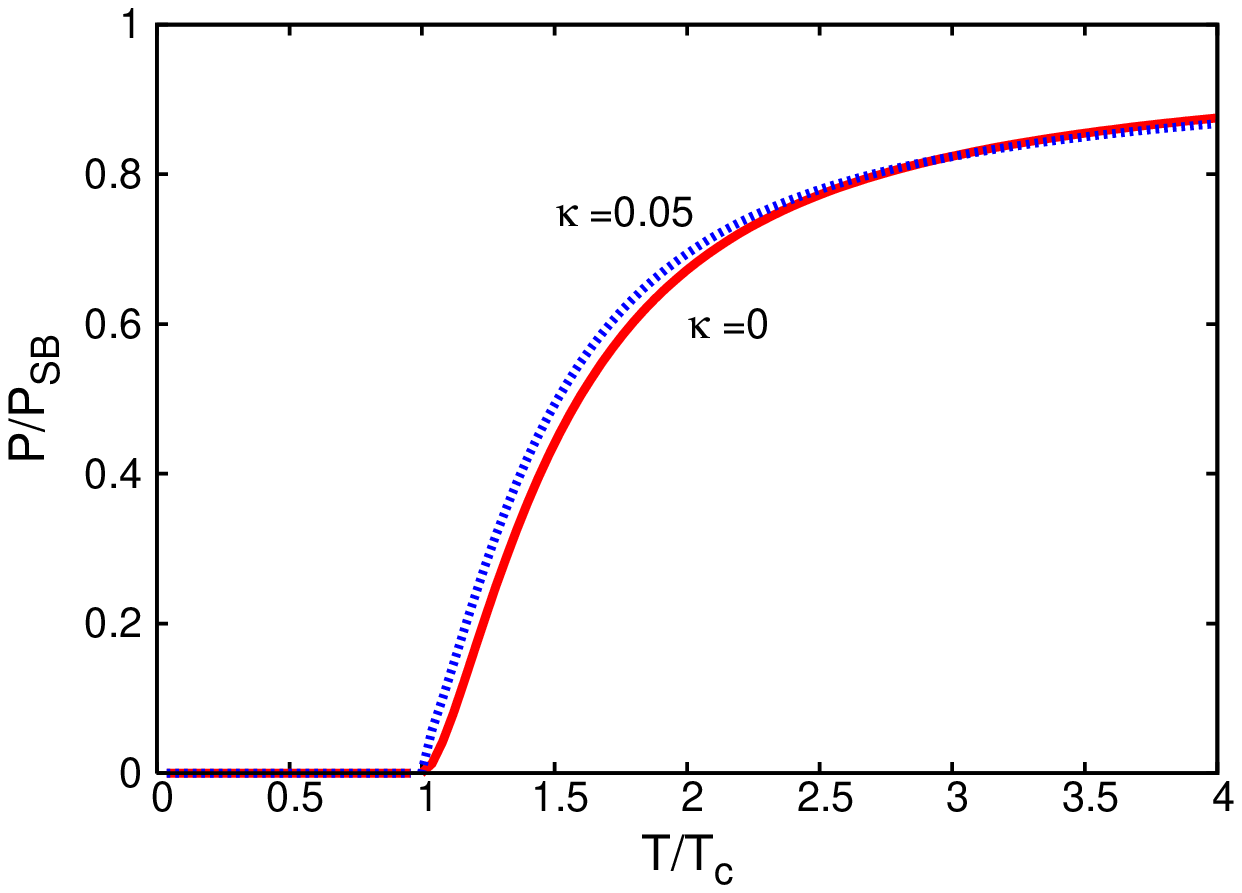}}
}
   \caption{$\Phi$ and $P/P_{SB}$ for $\kappa = 0$($T_0 = 0.27$ GeV)
 and $\kappa=0.5$($T_0=0.2555$ GeV). The value of $T_c$ is 0.270 GeV.
   }
\label{fg.purek}
\end{figure}

\noindent
To make a saddle point approximation to the mean fields, the 
potential density $\cU$ was
minimized w.r.t. $\Phi$ and $\bar{\Phi}$ in Ref. \cite{pnjl2}.
These were then used to obtain pressure $P=-\cU$. The coefficients 
$a_i$ ($i$=0,1,2,3) and $b_j$ ($j$=2,3,4) were fitted from Lattice 
data of pressure in pure gauge theory, and $T_0$ is precisely the 
transition temperature $T_c = 270$ MeV \cite{tcpg1,tcpg2,tcpg3}. 
As $T \rightarrow \infty$, $P/T^4 \rightarrow 16\pi^2/90$.
However, to take care of the effect of the Jacobian as discussed above,
we now propose to minimize the following modified potential,

\beqa
\frac{\cU^{\prime}(\Phi,\bar{\Phi})}{T^4} = 
\frac{\cU(\Phi,\bar{\Phi})}{T^4} - \kappa \ln [J(\Phi,\bar{\Phi})],
\label{eq.uup}
\eeqa

\noindent
where $\kappa$ is a dimensionless parameter to be determined
phenomenologically. The mean field value of
pressure is still obtained from the relation $P=-\cU$. A very 
simple example of this approach is demonstrated in the appendix.
Note that the Jacobian term is considered as an extra effective
term in the modified potential density implying a sort of 
normalized volume factor. This is quite natural as the form of
Eqn. \ref{eq.intmes} implies that there is a Jacobian sitting
at each and every space-time coordinate, depending on the value of 
$\Phi$ and $\bar{\Phi}$. 

\begin{figure}[!tbh]
\subfigure{
   {\includegraphics [scale=0.6] {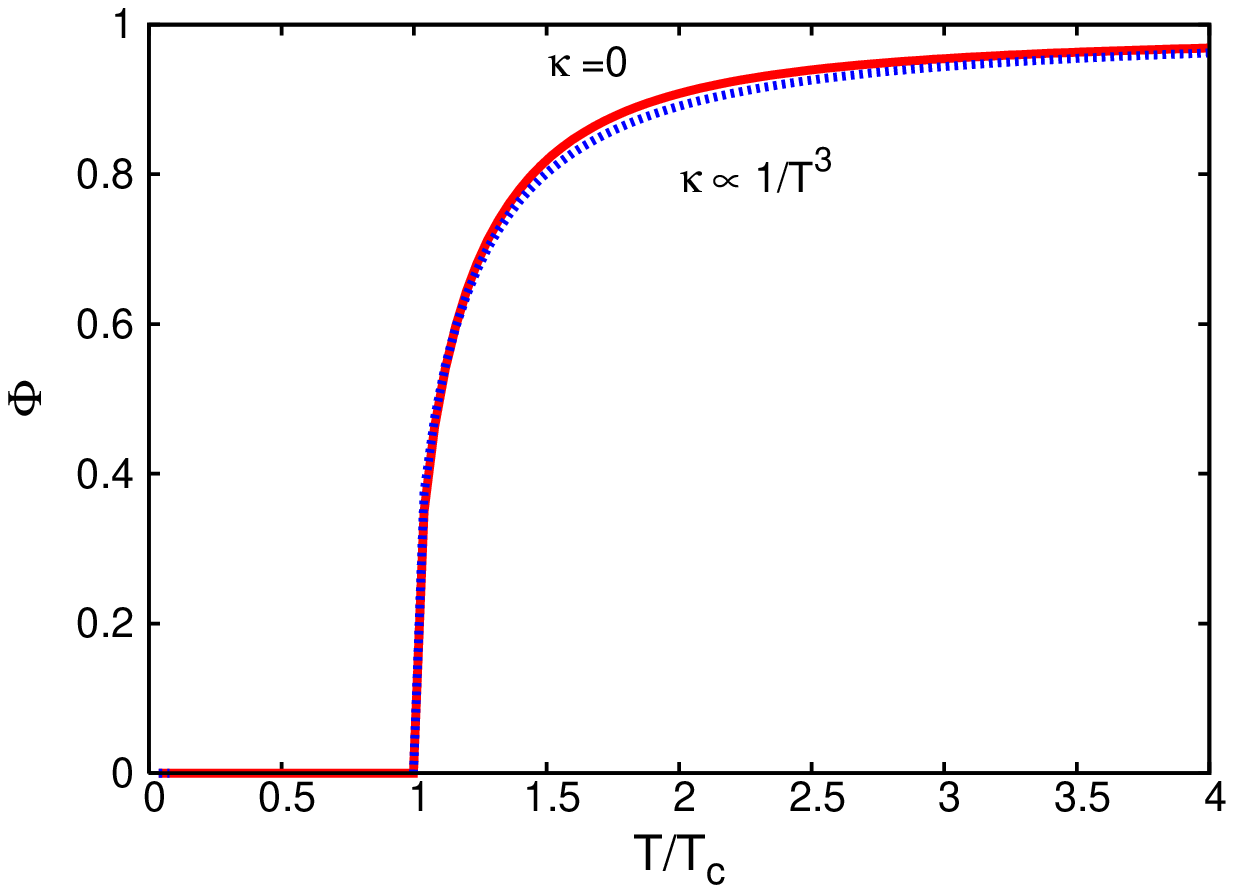}}
}
\hskip 0.15 in
\subfigure{
   {\includegraphics[scale=0.6]{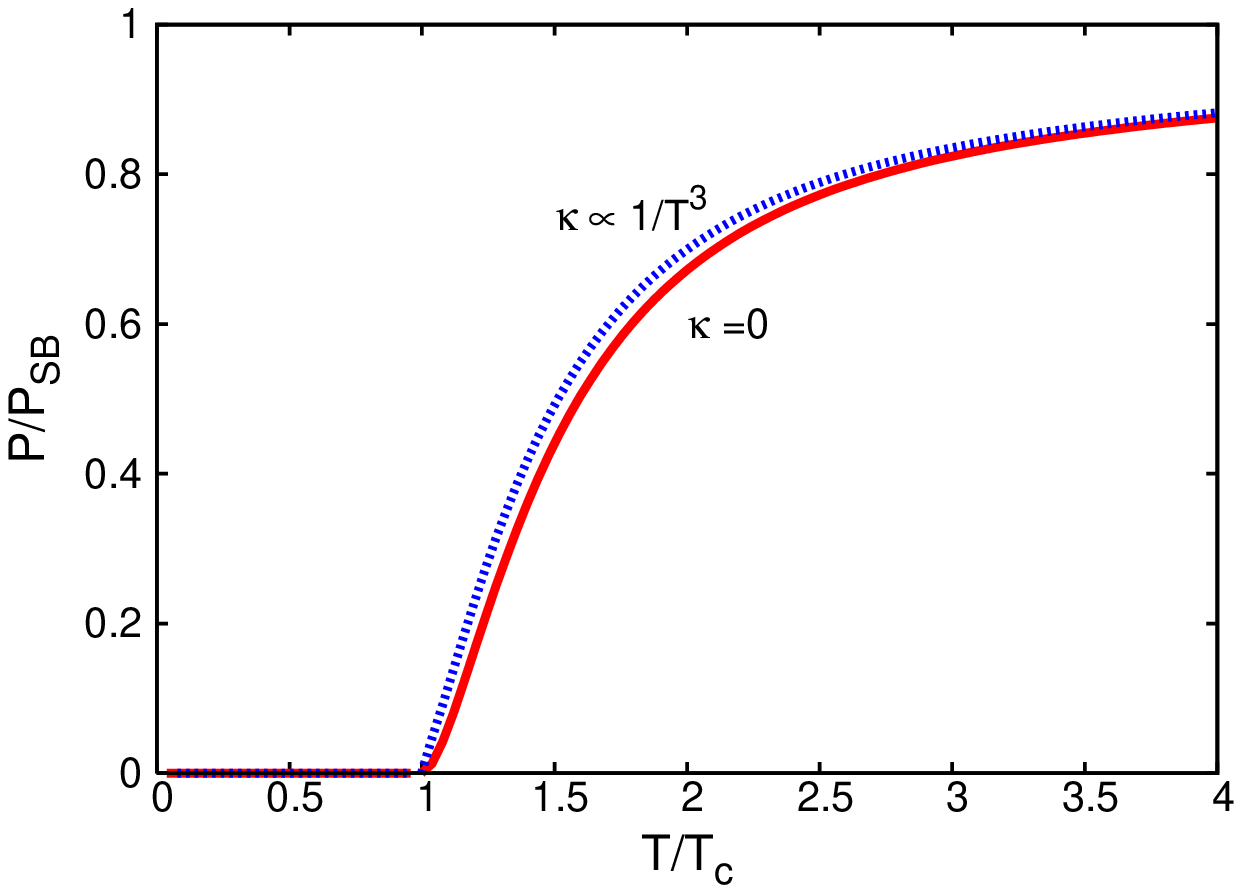}}
}
   \caption{$\Phi$ and $P/P_{SB}$ for $\kappa = 0$($T_0 = 0.27$ GeV)
 and $\kappa=(0.22\,T_0^3/T^3)$($T_0=0.2555$ GeV). Here $T_c =$ 
 0.270 GeV.
   }
\label{fg.purek3}
\end{figure}

With the new minimization condition all the coefficients should 
be estimated afresh. Instead, we retain the values of $a_i$ 
and $b_j$ obtained in \cite{pnjl2} and tune only the values of $T_0$ 
and $\kappa$. This is equivalent to
a correlated modification of the $a_i$ and $b_j$
keeping $T_0$ fixed at 270 MeV. 

We show the variation of the Polyakov loop and the pressure $P$ 
normalized to Stefan-Boltzmann (SB) pressure $P_{SB}$ for pure gauge
theory, as a function of 
temperature. In Fig.\ref{fg.purek} we have used a small non-zero 
constant value of $\kappa = 0.05$. In Fig.\ref{fg.purek3} we find
similar behaviour for a temperature dependent $\kappa = 0.22 T_0^3/T^3$.
In both the figures the $\kappa=0$ curves are for the Polyakov loop
model without the VdM term. Thus the parameter space of $\kappa$
is quite open at this stage.

Within the range of temperatures ($T < 3 T_c$) where the Polyakov 
loop model is supposed to be a good description of the system, 
our approach and that of Ref. \cite{ratti4} give similar results.
The reason behind this is that one can suitably adjust the parameters 
in both approaches. However, our method for introducing the VdM 
potential as discussed above, is very much different from that of 
Ref. \cite{ratti4}. The main difference is that the pressure computed
in Ref. \cite{ratti4} includes the VdM term. Thus the coefficient 
of the VdM term requires an inverse temperature dependence, so that 
on a naive extrapolation to high temperatures, the pressure does not 
blow up with the logarithm of the Jacobian. In that case
another problem crops up with the remaining part of the thermodynamic 
potential, which at high temperatures has no bound, contrary to the 
claim that $\Phi \rightarrow 1$ as $T \rightarrow \infty$. Precisely 
because $\Phi$ should go to 1 as $T \rightarrow \infty$ we believe 
that the VdM term should be very important at high temperatures to
constrain the maximum value of $\Phi$ to 1. 

The exercise for introducing a VdM term for the Polyakov loop model
itself has nothing new to offer. Even without it the potential $\cU$
was able to describe the pure glue theory quite well. However its
importance becomes evident in the PNJL model. The Polyakov loop has
a coupling to the fermionic part as will be seen in the corresponding 
thermodynamic potential below, which forces the $\Phi$
to be greater than 1, and more so as the chemical potential is 
increased. The VdM term can inhibit such a behaviour.

The thermodynamic potential of the PNJL model \cite{pnjl2,pnjl3,pnjl4}
is given as,

\beqa
\Omega&=&{\cal U}\left(\Phi,\bar{\Phi},T\right)+
2 G_1(\sigma_u^2 + \sigma_d^2) + 4 G_2 \sigma_u \sigma_d 
\nonumber \\
&-& \sum_{f=u,d}
2\,T\int\frac{\mathrm{d}^3p}{\left(2\pi\right)^3}
\left\{ \ln\left[1+3\left(\Phi+\bar{\Phi}\mathrm{e}^
{-\left(E_f-\mu_f\right)/T}\right)\mathrm{e}^{-\left(E_f-\mu_f\right)/T}
 + \mathrm{e}^{-3\left(E_f-\mu_f\right)/T}\right]\right . \nonumber\\
&+& 
\left . 
\ln\left[1+3\left(\bar{\Phi}+\Phi\mathrm{e}^{-\left(E_f+\mu_f\right)/T}
\right)\mathrm{e}^{-\left(E_f+\mu_f\right)/T}+
\mathrm{e}^{-3\left(E_f+\mu_f\right)/T}\right] \right\}
- \sum_{f=u,d} 6\int\frac{\mathrm{d}^3p}{\left(2\pi\right)^3}{E_f}
\theta\left(\Lambda^2-\vec{p}^{~2}\right) ~~~.
\label{omega}
\eeqa
Here quark condensates for the two light flavors $u$ and $d$ are given by 
$\sigma_u = <\bar{u}u>$ and $\sigma_d=<\bar{d}d>$ respectively, 
and the respective
chemical potentials are $\mu_u$ and $\mu_d$. Note that 
$\mu_0 = (\mu_u+\mu_d)/2$ and $\mu_I = (\mu_u-\mu_d)/2$. The quasi-particle 
energies are $E_{u,d}=\sqrt{\vec{p}^{~2}+m_{u,d}^2}$, where 
$m_{u,d}=m_0-4 G_1 \sigma_{u,d} -4 G_2 \sigma_{d,u}$ are the 
constituent quark masses and $m_0$ is the current quark mass
(we assume flavour degeneracy).  $G_1$ and $G_2$ are the effective 
coupling strengths of a local, chiral symmetric four-point 
interaction. We take $G_1 = G_2 = G/4$, where $G$ is the coupling 
used in Ref. \cite{pnjl2}. 
$\Lambda$ is the 3-momentum cutoff in the NJL model.
${\cal U}\left(\Phi,\bar{\Phi},T\right)$ is the effective
potential for $\Phi$ and $\bar{\Phi}$ as given in Eqn. \ref{eq.uu}.
We locate the transition temperature in this model from the peaks 
in the temperature variation of $d\Phi/dT$ and $d\sigma_{u,d}/dT$.

Similar to the case of the Polyakov loop model we would now obtain the
mean fields by minimizing,
\beqa
\frac{\Omega^{\prime}}{T^4} = \frac{\Omega}{T^4} - 
\kappa\,\ln [J(\Phi,\bar{\Phi}]
\label{eq.omegap}
\eeqa

The coefficient $\kappa$ in the VdM term can in general have some 
temperature and/or chemical potential dependence. Here we take
a constant value $\kappa = 0.2$ which suffices for the purpose of the
present work. To set this value we looked at the two important 
quantities affected by the VdM term. First one is $\Phi$ which
decreases with the increase of $\kappa$ and hence decreases the pressure. 
Second one is the transition temperature which increases with $\kappa$.
Thus we try to optimize $\kappa$ to get both the pressure and
the transition temperature as close as possible to the LQCD results
for two quark flavours. 

On a naive extrapolation of this model to large chemical potentials,
the $\Phi$ and $\bar{\Phi}$ should grow towards 1 (deconfinement at 
large chemical potential) even at very low temperatures.
Thus again the logarithmic term blows up. So if pressure is 
computed including the VdM term as is done in Ref. \cite{ratti4},
an anomalous logarithmic divergence would come up. There may be some
new physics that can obscure such terms by making $\kappa \rightarrow 0$
as $\mu \rightarrow \infty$. But that would again run into a problem
in restricting $\Phi$ in the domain $0 \le \Phi \le 1$.

Apart from the difference in the treatment of the VdM term we would
now remove the condition $\Phi = \bar{\Phi}$ used in \cite{ratti4}, 
since it has important implications for susceptibilities. 

Before going over to our results let us take a digression to the
Lattice computation of $\Phi$. On the lattice $\Phi$ is computed
from the relation \cite{polyl},

\beqa
\Phi(T) = \exp\,(-\bigtriangleup F_{q \bar{q}}(\infty,T)/2T),
\label{eq.screen}
\eeqa

\noindent
where, $\bigtriangleup F_{q \bar{q}}(\infty,T) = 
F_{q \bar{q}}(\infty,T) - F_{00}(T)$, and
$F_{q \bar{q}}(r,T)$ is the free energy of a pair of heavy quark and 
anti-quark at a separation $r$ at a temperature $T$. This has been
used to define a renormalized Polyakov loop in lattice simulations of
both pure gluon \cite{polatgl1,polatgl2} and full QCD \cite{polatfm}.
In fact the data of \cite{polatgl1} was used to obtain the different 
parameters of the Polyakov
loop model in \cite{pnjl2}, and is being used by us here, and in that
sense $\Phi$ is the renormalized Polyakov loop. But even in this
exercise the $\Phi$ in the Polyakov loop model of \cite{pnjl2}
goes to 1 at large T and is thus different from lattice results for
$T > T_c$. On the lattice the value of $\Phi$ goes above 1 for $T > T_c$.
It has been argued
that since the $\Phi$ measured in lattice simulations is a renormalized
quantity, it is no more a character of the group SU(3) and is
thus not limited to values below 1. From Eqn.\ref{eq.screen},
it is evident that $\Phi > 1$ only when 
$\bigtriangleup F_{q \bar{q}}(\infty,T) < 0$, and this can be very
easily seen to be true in the lattice simulations and happens for 
$T > T_c$. Now, the free energy $F_{q \bar{q}}(r,T)$ can be considered 
to be composed
of three components, namely, a confining potential, a screening 
potential and an entropy part. For low temperatures the confining
part is dominant and $\bigtriangleup F_{q \bar{q}}(\infty,T) > 0$.
In the deconfined phase for large distances, the screening potential
drops out so the entropy part is dominant which could lead to
$\bigtriangleup F_{q \bar{q}}(\infty,T) \simeq 
- T \bigtriangleup S_{q \bar{q}}(T) < 0$, where 
$\bigtriangleup S_{q \bar{q}}(T) = S_{q \bar{q}}(T) - S_{00}$,
and $S_{q \bar{q}}(T)$ denotes the entropy of the system with
a pair of quark and anti-quark. However the heavy quarks as
such are not expected to contribute significantly to the entropy and
it seems natural to have $\bigtriangleup S_{q \bar{q}}(T) = 0$, and
thus $\bigtriangleup F_{q \bar{q}}(\infty,T) = 0$ for $T > T_c$. 
Instead the value is negative on the lattice and
$\bigtriangleup F_{q \bar{q}}(\infty,T) \rightarrow - \infty$ as
$T \rightarrow \infty$, leading to $\Phi \rightarrow \infty$. 
One has to then worry about what can bend it down towards 1 
at asymptotic temperatures as was observed by Gava and Jengo in 
perturbative 
evaluation of $\Phi$ \cite{gava}. However this perturbative
calculation also points to the fact that as the temperature is
lowered from asymptotic values the $\Phi$ is greater than 1. Also 
recent continuum estimates in chiral quark models \cite{enrique1} 
using dimensional reduction find close agreement with both lattice 
and perturbative calculations.

On the other hand another lattice computation of the Polyakov 
loop in pure glue theory uses a renormalization dependent on temperature
instead on the lattice spacing and finds the values to remain below
1 at least upto $T \sim 3.5 T_c$ \cite{dumitrupol}. We thus admit that 
the state of affairs with the lattice computation of $\Phi$ is not
very clear to us at this stage. There is a missing link
from quantum computations to our matrix model mean-field computations.

\section{Results and discussions} \label{sc.results}

\begin{figure}[!tbh]
\subfigure[]{
   {\includegraphics [scale=0.6] {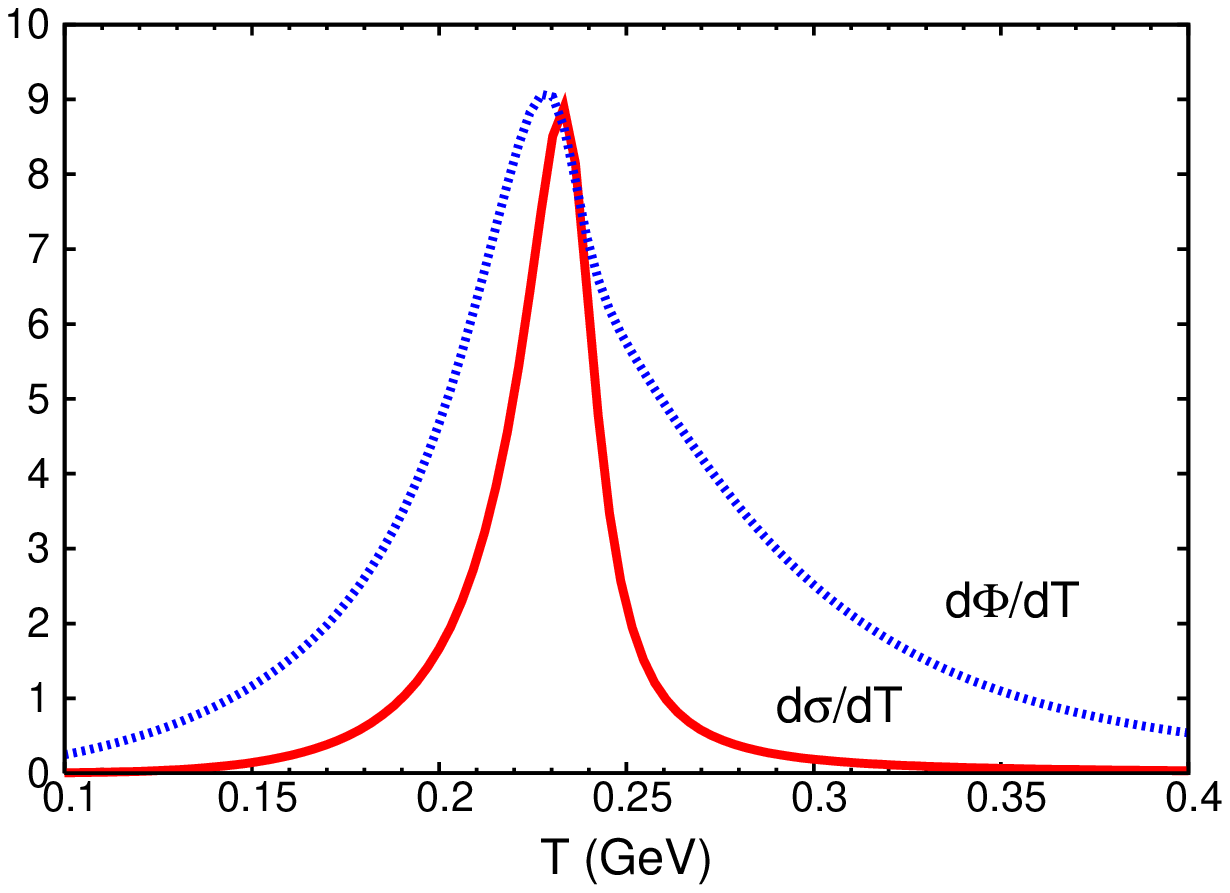}}
\label{fg.peaks}
}
\hskip 0.15 in
\subfigure[]{
   {\includegraphics[scale=0.6]{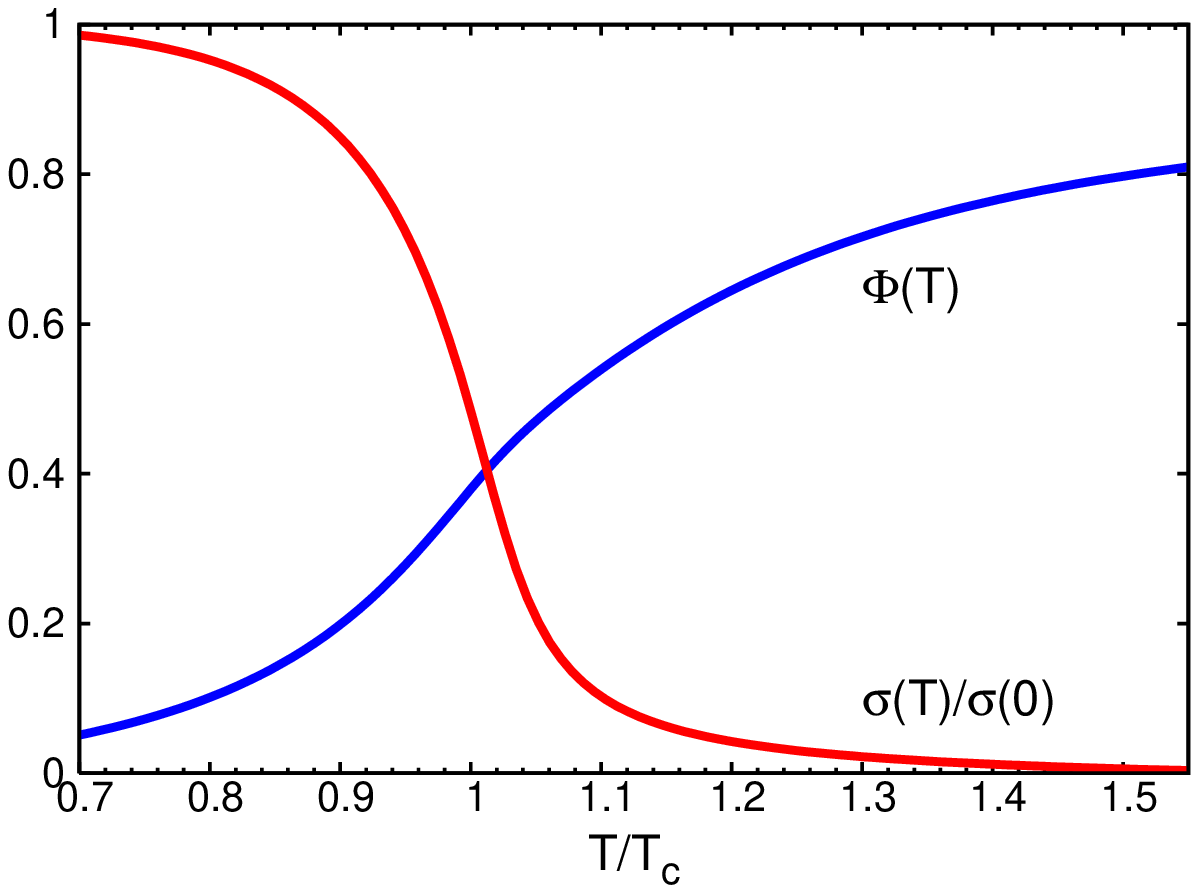}}
\label{fg.phisig}
}
\vskip 0.1 in
\subfigure[]{
   {\includegraphics [scale=0.6] {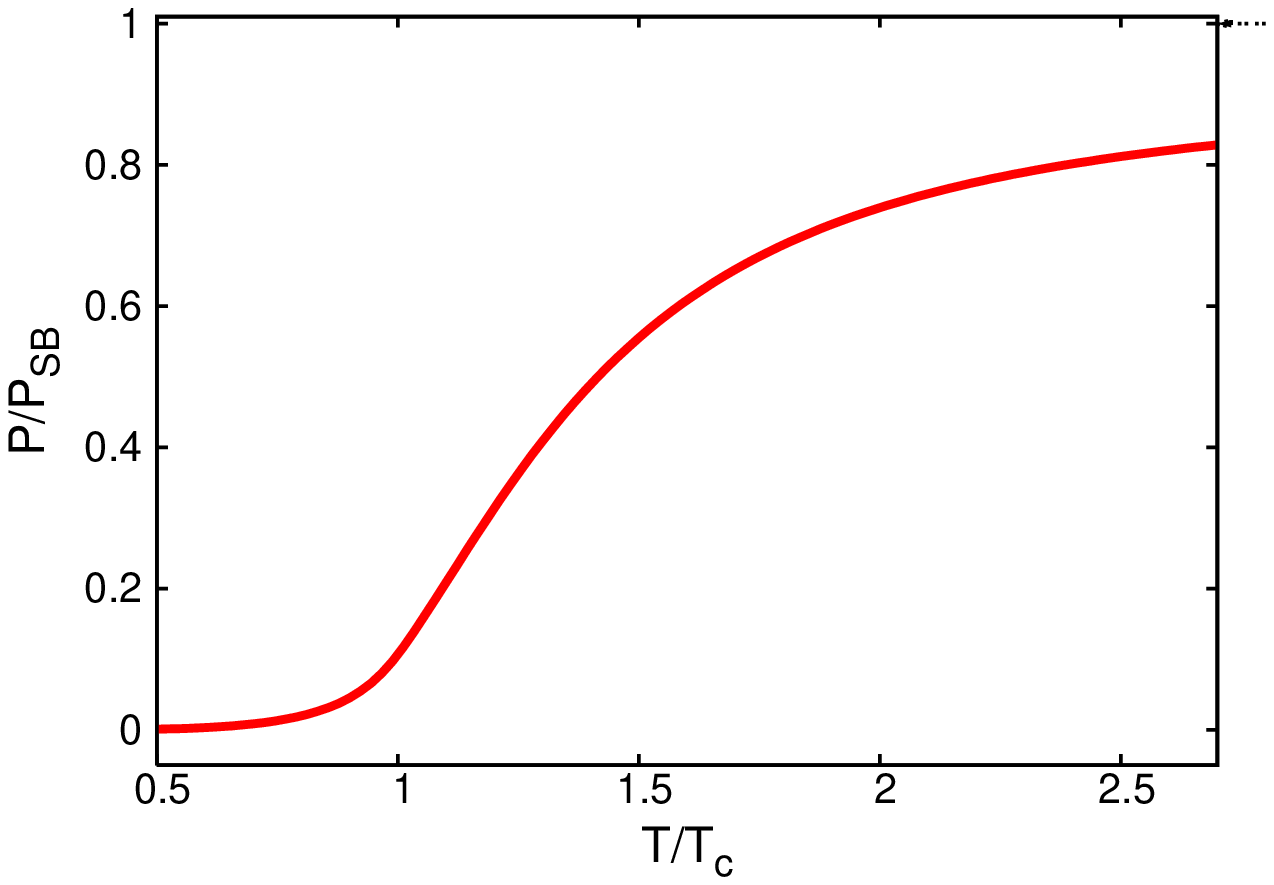}}
\label{fg.presr}
}
   \caption{(a): Peaks in ${\rm d}\Phi /{\rm d}T$ and
   ${\rm d}\sigma /{\rm d}T$ sets the $T_c$ at around 230 MeV.
   (b): $\Phi$ and $\sigma$ as functions of $T/T_c$.\\
   {\it Note:} In this figure $\sigma = G(\sigma_u + \sigma_d)$.
   }
\end{figure}

\subsection{PNJL Model: Pressure, specific heat and speed of sound}
\label{sc.pressure}

Now we discuss the results for the PNJL model with VdM term. Here the 
$\Omega^{\prime}$ as given in Eqn. \ref{eq.omegap} is minimized with 
respect to the fields and all the thermodynamic quantities are obtained 
using these values. The peaks of the $d\Phi/dT$ and $d\sigma_{u,d}/dT$ 
curves, as shown in Fig. \ref{fg.peaks}, differs by 5 MeV.
Their average position, which is at 230 MeV, is taken as the transition 
(or crossover) temperature $T_c$. In spite of the significant difference 
of $T_c$ in the PNJL model with the corresponding LQCD value of 
192(7)(4) MeV \cite{cheng}, the thermodynamic quantities when 
plotted against the scaled temperature $T/T_c$ show similar behaviour.
We shall henceforth show the temperature dependences in terms of $T/T_c$. 

As mentioned earlier we are using an optimized value of $\kappa = 0.2$. 
The temperature dependence of the fields are shown in Fig. 
\ref{fg.phisig}. It agrees reasonably with that of the LQCD results 
as shown in Fig. 1 of Ref. \cite{petcy}. The scaled pressure $P/P_{SB}$
is plotted in Fig. \ref{fg.presr}. It slightly overestimates the
LQCD pressure \cite{leos}. However it agrees well with the recent 
LQCD results for 2+1 flavors with almost physical quark masses 
\cite{datta}.

Now, the energy density $\epsilon$ is obtained from the relation,

\beqa
\epsilon 
   = - T^2 \left . \frac{\partial}{\partial T} 
        \left(\frac{\Omega}{T}\right) \right |_V
   = - T \left . {\frac{\partial \Omega}{\partial T}} \right |_V 
              + \Omega ~~~.
\eeqa

\noindent
The rate of change of energy density $\epsilon$ with temperature 
at constant volume is the specific heat $C_V$ which is given as,

\beqa
C_V = \left . {\frac{\partial \epsilon}{\partial T}} \right |_V 
    = - \left . T {\frac{\partial^2 \Omega}{\partial T^2}} \right |_V ~~~.
\label{sph}
\eeqa

\noindent
The square of velocity of sound at constant entropy $S$ is given by,

\beqa
v_s^2 = \left . {\frac{\partial P}{\partial \epsilon}} \right |_S 
      = \left . {\frac{\partial P}{\partial T}} \right |_V \left /
        \left . {\frac{\partial \epsilon}{\partial T}} \right |_V \right .
      = \left . {\frac{\partial \Omega}{\partial T}} \right |_V \left /
        \left . T {\frac{\partial^2 \Omega}{\partial T^2}} \right |_V 
        \right .  ~~~.
\label{sps}
\eeqa

\noindent
The conformal measure is given by,

\beqa
\cC=\Delta/\epsilon \qquad ; \qquad \Delta = \epsilon - 3P
\label{cnm}
\eeqa

\begin{figure}[!tbh]
\subfigure[]{
   {\includegraphics [scale=0.6] {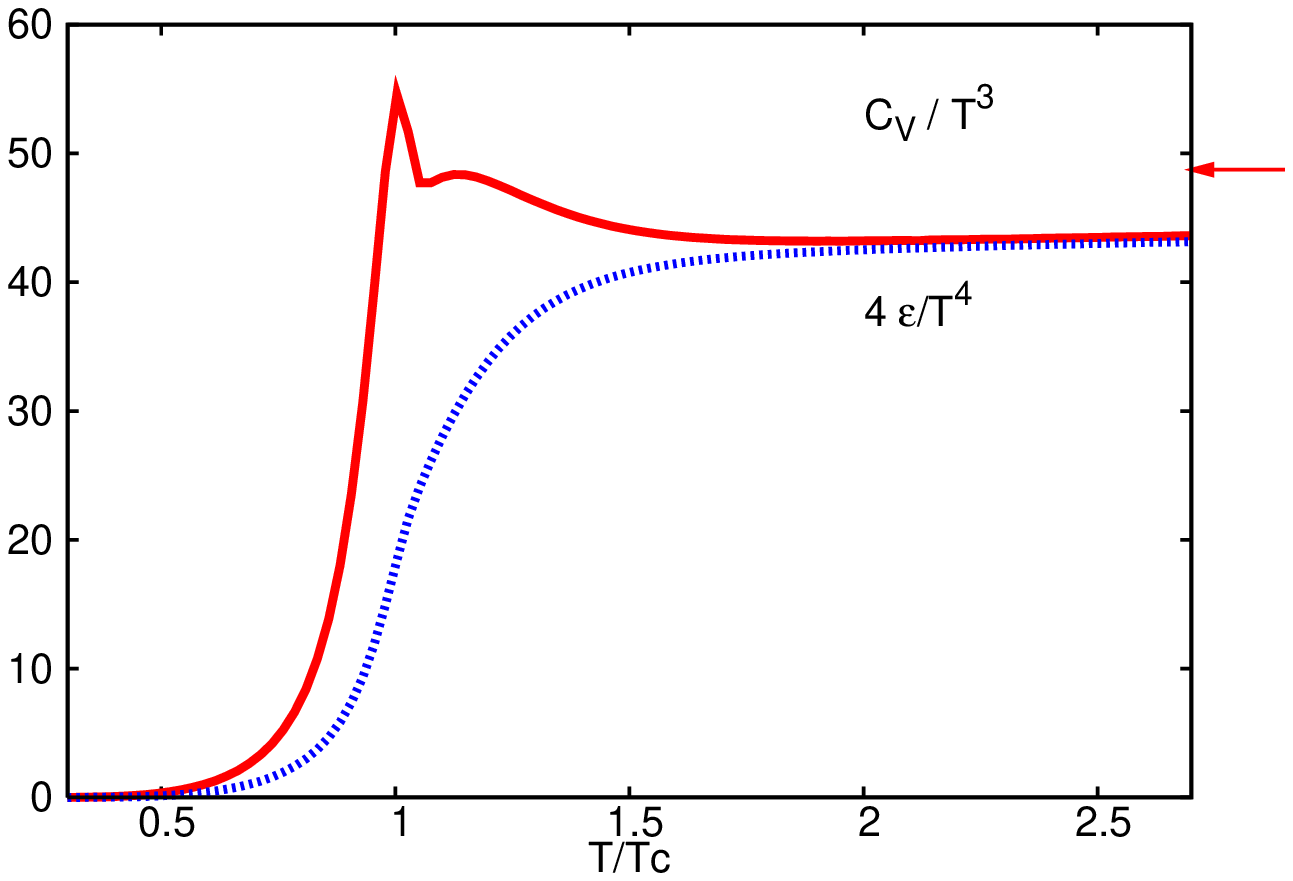}}
\label{fg.cv}
}
\hskip 0.15 in
\subfigure[]{
   {\includegraphics[scale=0.6]{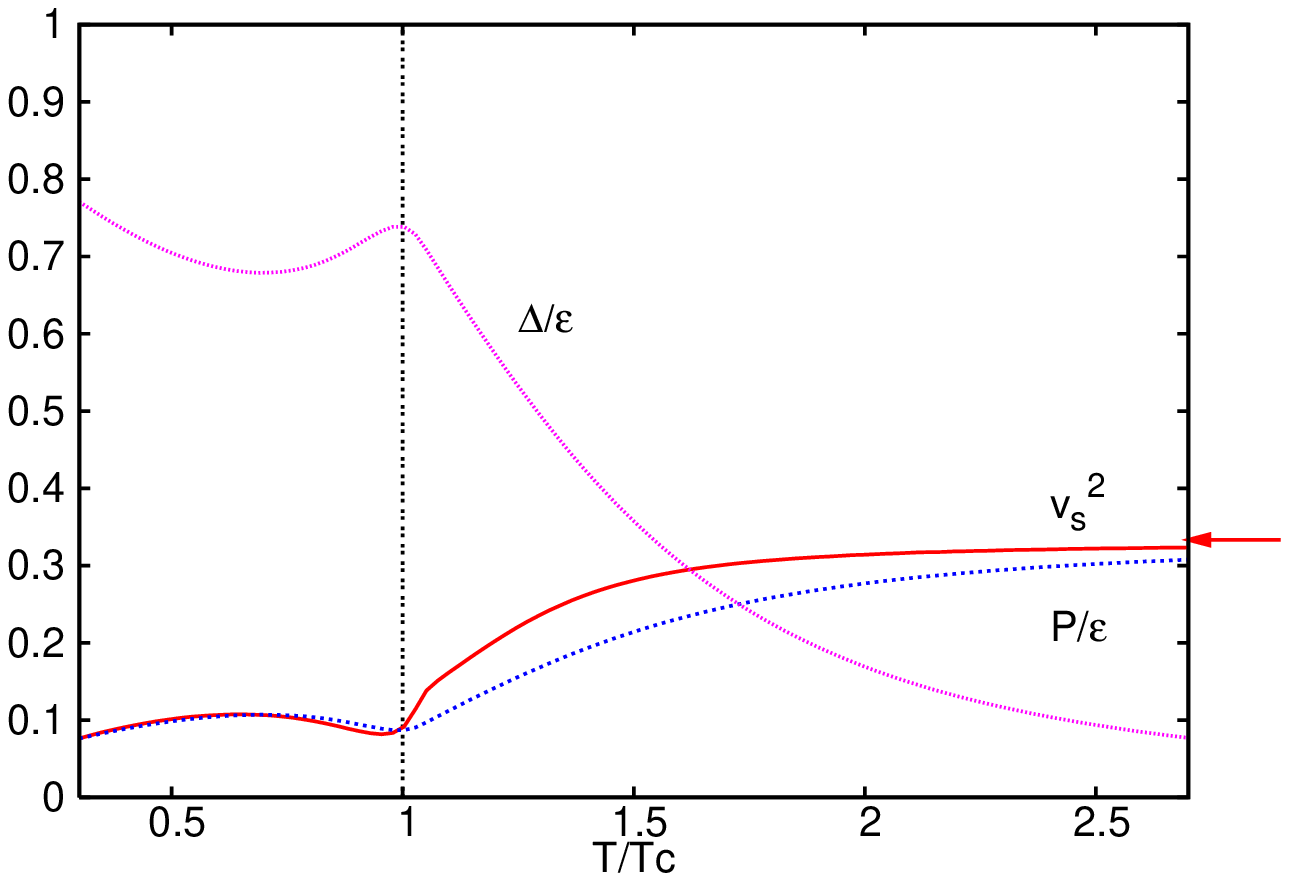}}
\label{fg.cs}
}
   \caption{(a): Temperature dependence of energy density $\epsilon$ 
            and specific heat $C_V$. (b): Temperature dependence of
            squared speed of sound $v_s^2$ and conformal measure
            $\Delta/\epsilon$. The arrows on the right show the
            corresponding SB limit.
	   }
\label{fg.cvcs}
\end{figure}

These quantities are plotted in Fig. \ref{fg.cvcs}. At higher
temperatures the $C_V$ is slightly lower than the values obtained
in \cite{pnjl3}. However, the velocity of sound and the conformal
measure remain unaltered in the whole range of temperatures. Thus
the VdM term affects $C_V$ but not quantities involving ratios
of pressure and energy density e.g. $v_s^2$ and $\cC$. It is 
interesting to note that our earlier \cite{pnjl3} as well as the 
present work, have been able to predict the value of $v_s^2$ quite
well when compared to the recent LQCD results \cite{datta}.
We hope similar encouraging results would be obtained on the lattice
for the specific heat.

\subsection{Taylor expansion of Pressure}
\label{sc.tayexp}

\begin{figure}[!tbh]
\subfigure[]{
\label{fg.c2}
   {\includegraphics [scale=0.6] {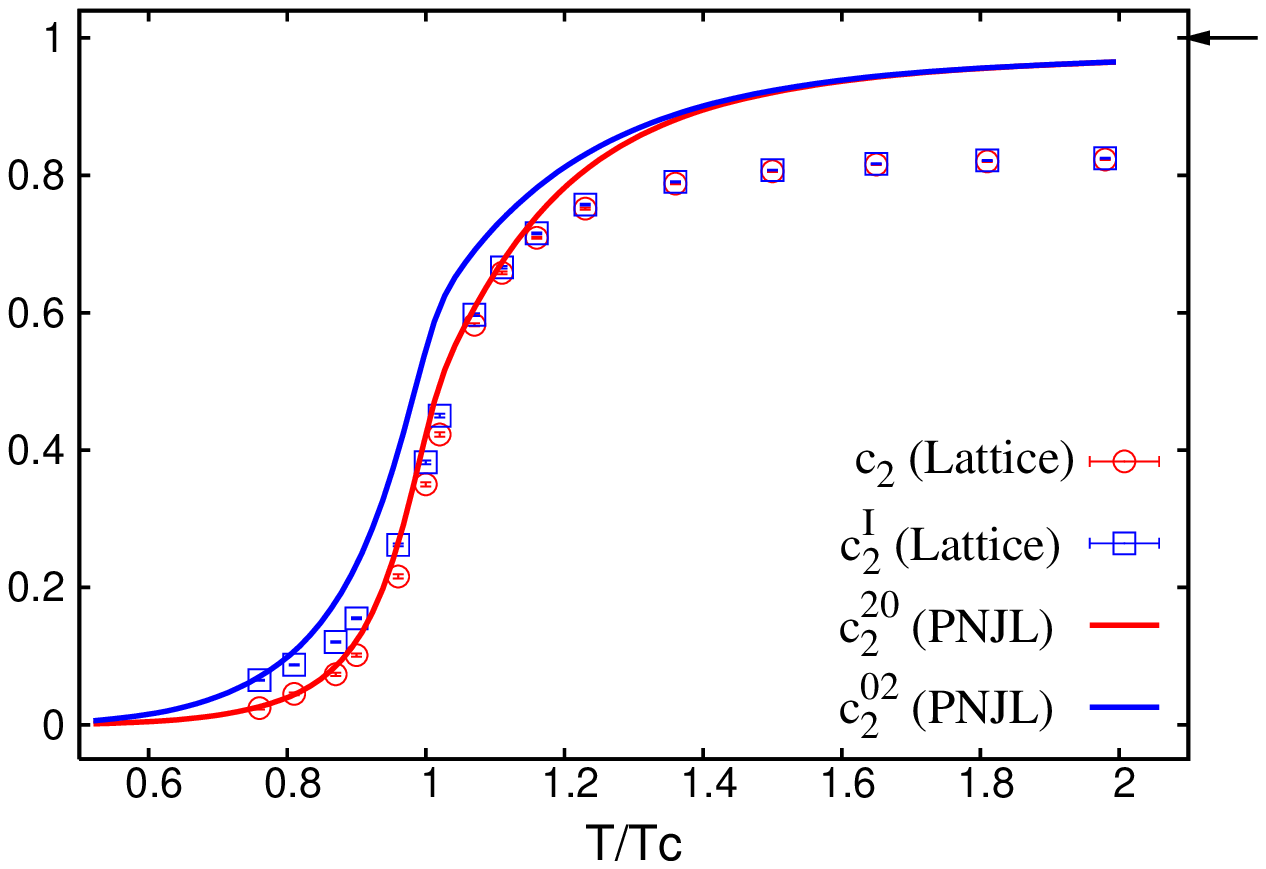}}
}
\hskip 0.1 in
\subfigure[]{
\label{fg.c4}
   {\includegraphics [scale=0.6] {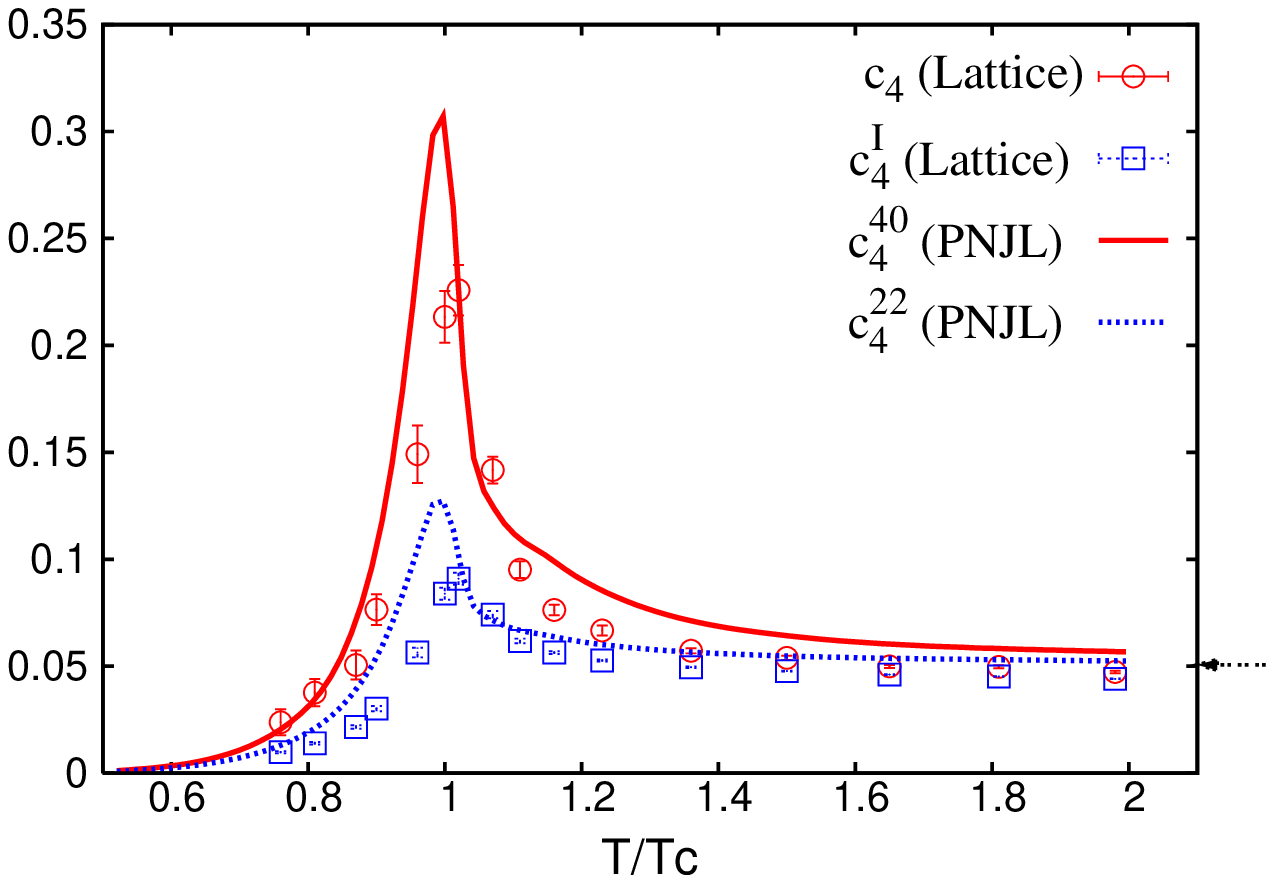}}
}
\vskip 0.1 in
\subfigure[]{
\label{fg.c6}
   {\includegraphics [scale=0.6] {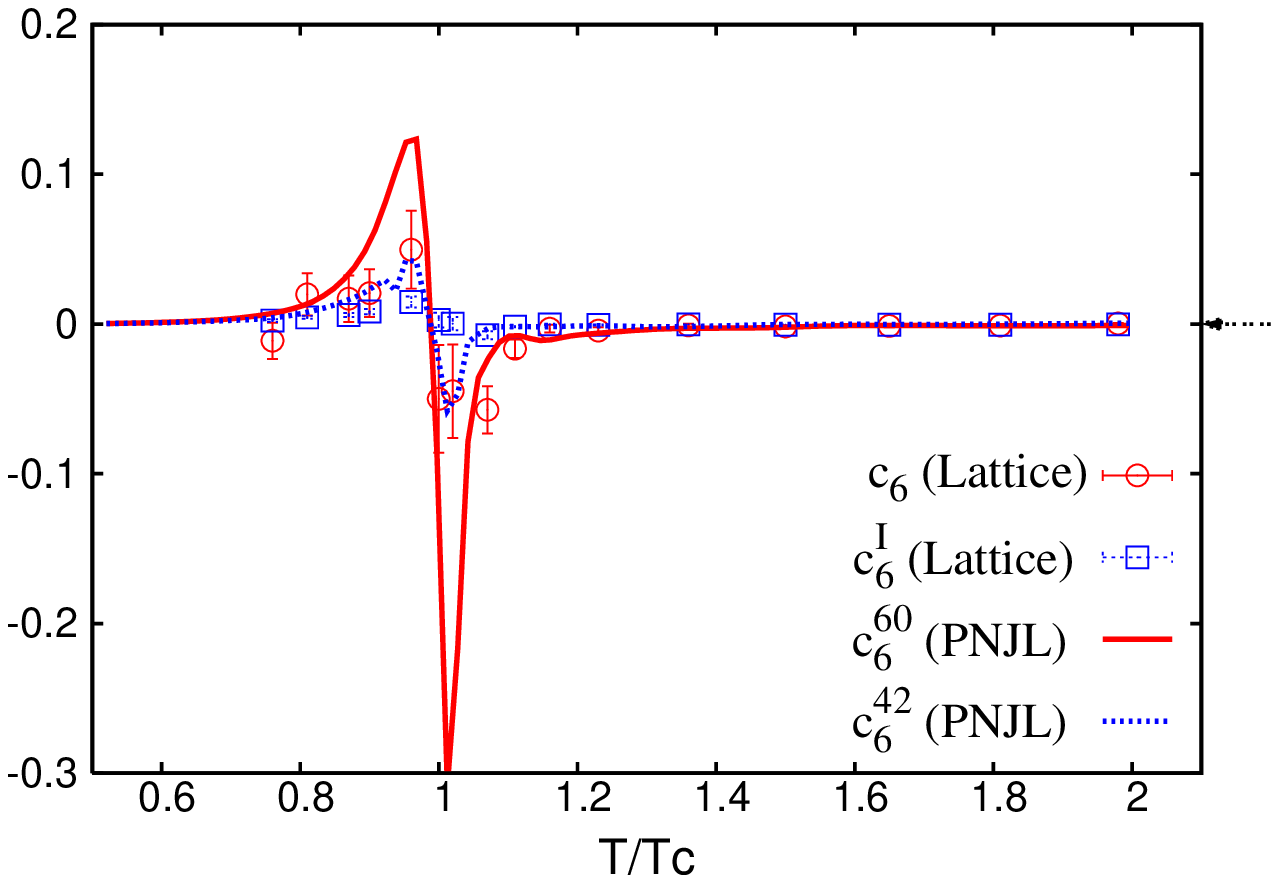}}
}
   \caption{The Taylor expansion coefficients of pressure in
            quark number and isospin chemical potentials as functions 
            of $T/T_c$. Symbols are LQCD data \cite{sixx}. Arrows on 
            the right indicate the corresponding ideal gas values.
      }
\label{fg.scnord}\end{figure}

The Taylor expansion coefficients of pressure with respect to
chemical potentials have been the focus of comparison of PNJL and
LQCD results \cite{pnjl3,pnjl4,ratti4,ratti5}. Here we have 
expanded the scaled pressure ($P/T^4$) in a Taylor series for the 
quark number and isospin number chemical potentials, 
$\mu_0$ and $\mu_I$ respectively,

\beqa
\frac{P(T,\mu_0,\mu_I)}{T^4}= 
\sum^{\infty}_{n=0} \sum^{n}_{j=0} \frac{n !}{j ! (n-j) !} c^{jk}_n(T)
\left(\frac{\mu_0}{T}\right)^j \left(\frac{\mu_I}{T}\right)^k ~~~;~ k=n-j,
\label{tay}
\eeqa
where,
\beqa
c^{jk}_n(T) = 
\frac{1}{n!} \frac{\partial^n \left ({P(T,\mu_0,\mu_I) / T^4} \right )}
{\partial \left(\frac{\mu_0}{T}\right)^j \partial \left(\frac{\mu_I}{T}\right)^k}
\Big|_{\mu_0=0,\mu_I=0} ~~~.
\label{taycoff}
\eeqa
The $n= {\rm odd}$ terms vanish due to CP symmetry. Even for the 
$n= {\rm even}$ terms, due to flavour degeneracy all the coefficients 
$c^{jk}_n$ with $j$ and $k$ both odd vanish identically.
We evaluate all the 10 nonzero coefficients (including
the pressure at $\mu_0 = \mu_I = 0$) upto order $n=6$ and compare them 
to LQCD data. These coefficients were evaluated in \cite{pnjl3,pnjl4}
and certain differences were found w.r.t. LQCD data. We shall now 
discuss the effects of the VdM term on these coefficients.

The coefficients we deal with are given by,

\beqa
c_n(T) &=& \frac{1}{n!} \left. \frac{\partial^n \left ({P(T,\mu_0) / T^4}
\right ) }
{\partial \left(\frac{\mu_0}{T}\right)^n}\right|_{\mu_0=0} = c^{n0}_n~~~,
\eeqa

\beqa
c^I_n(T) &=& \left. {\frac{1}{n!} \frac{\partial^n \left ({P(T,\mu_0,\mu_I) /
T^4} \right ) }{
\partial \left(\frac{\mu_0 }{ T }\right)^{n-2}
\partial \left(\frac{\mu_I }{T }\right)^2 
}}\right|_{\mu_0=0,\mu_I=0} = c^{(n-2) 2}_n~~~; ~ n > 1.
\eeqa

We present the QNS, INS and their higher order derivatives with respect
to $\mu_0$ in Fig.\ \ref{fg.scnord}. We have plotted the LQCD data
from Ref.\ \cite{sixx} for quantitative comparison.  At the second order
(Fig. \ref{fg.c2}) we find that the QNS $c_2$ compares well 
with the LQCD data upto about 1.2 $T_c$. Thereafter the
PNJL values rise up towards the SB limit, while the
LQCD values saturate at about $80\%$ of this limit. The INS $c^I_2$ 
also shows similar behaviour, but at lower temperatures it goes
slightly above the corresponding LQCD values. There is no significant
difference of $c^I_2$ with and without the VdM term. However $c_2$ was 
close to the LQCD result without VdM term \cite{pnjl4}, but now at
high temperatures it goes above the LQCD values and approaches $c^I_2$. 
Thus at high temperatures these coefficients overestimate the LQCD 
results but both are almost equal to each other, similar to that 
observed on the Lattice. This was not so without the VdM term \cite{pnjl4}.

Now we discuss the $4^{th}$ order coefficients (Fig.\ref{fg.c4}). The 
values of $c_4$ 
in the PNJL model with VdM term matches closely with those of LQCD 
data for the full range of temperatures. This is in contrast to that
found without the VdM term \cite{pnjl3} where they were close
only upto $T \sim 1.1 T_c$. The VdM term does not affect the
coefficient $c^I_4$ which agrees well with LQCD data for the full 
range of $T$. Also both these coefficients approach each other as well
as the corresponding SB limit.  
At the $6^{th}$ order (Fig.\ref{fg.c6}) the coefficients do not seem 
to be affected by the VdM term. 

Thus we write down the salient features regarding the Taylor coefficients
in this modified PNJL model:

\begin{itemize}

\item All the coefficients start approaching their respective SB limit 
around $2 T_c$.

\item Both the QNS and INS approach each other at $2 T_c$. This is also
true for their corresponding responses to quark chemical potential given
by the $4^{th}$ and $6^{th}$ order coefficients.

\item At high temperatures, except $c_2$ and $c^I_2$, all the 
coefficients compare well quantitatively with the LQCD data.

\item The main effect of the VdM term is to move
$c_2$ and $c_4$ close to their respective SB limits.

\end{itemize}

\begin{figure}[!tbh]
\subfigure[]{
\label{fg.phmub}
   {\includegraphics [scale=0.6] {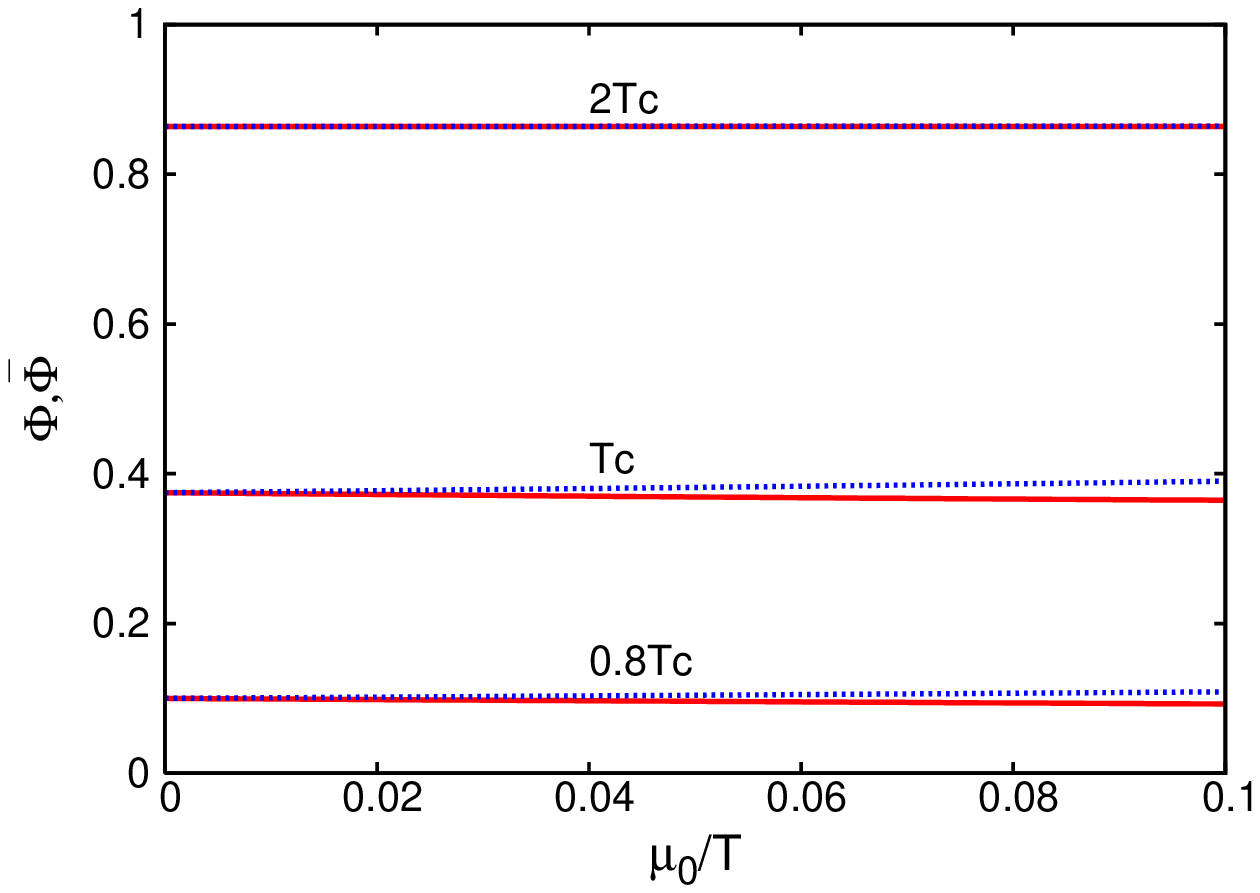}}
}
\hskip 0.15 in
\subfigure[]{
\label{fg.phmui}
   {\includegraphics[scale=0.6]{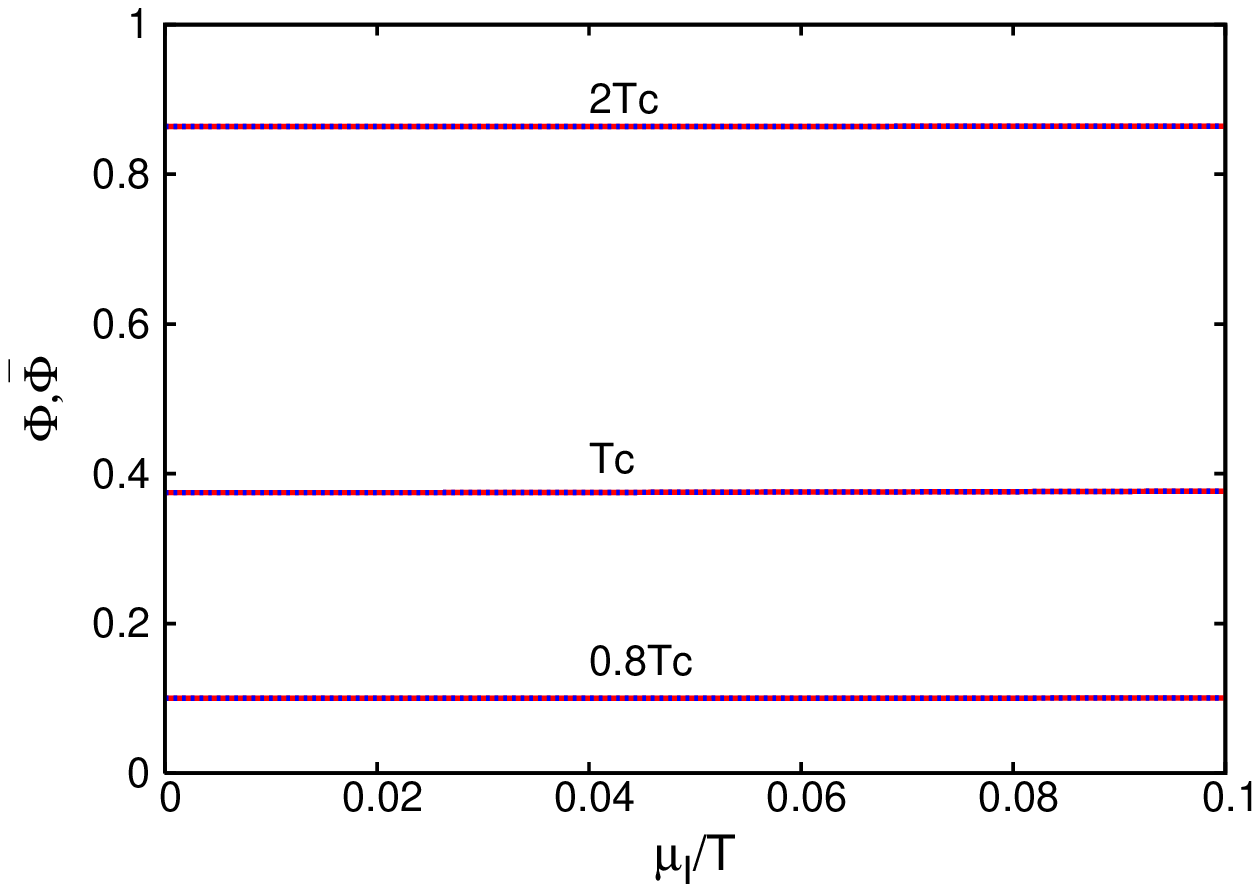}}
}
   \caption{(a): $\Phi$ (solid lines) decreases and
   $\bar{\Phi}$ (dotted lines) increases as a function of $\mu_0/T$
   ($\mu_I=0$) at low temperatures and almost equal and constant at
   high temperatures. (b): $\Phi$ (solid lines) and
   $\bar{\Phi}$ (dotted lines) are equal and almost constant as a
   function of $\mu_I/T$ ($\mu_0=0$). } 
\label{fg.phimub}\end{figure}

We have emphasized the role of the Polyakov loop in obtaining the
values of the Taylor coefficients in our earlier works \cite{pnjl3,pnjl4}.
In those works we found firstly that the Polyakov loop goes above 1 at
high temperatures and also has a significant dependence on $\mu_0$ but
not on $\mu_I$. Here as shown in Fig. \ref{fg.phimub}, the VdM term 
restricts the value of $\Phi$ within 1, and also the $\mu_0$ dependence 
at higher temperatures is almost negligible. Thus even 
the splitting between $\Phi$ and $\bar{\Phi}$ has
almost disappeared. We note here that though we let $\Phi$ and 
$\bar{\Phi}$ to be different, they come out to be almost equal at
high temperatures. This is in contrast to imposing $\Phi = \bar{\Phi}$
for the full range of temperatures as done in Ref. \cite{ratti4}. 
The difference between $\Phi$ and $\bar{\Phi}$ is responsible for the
difference of $c_2$ and $c^I_2$ in the intermediate temperatures. 

To complete the comparison with the LQCD data we have looked at the
flavour diagonal ($c_n^{uu}$) and flavour off-diagonal ($c_n^{ud}$)
susceptibilities defined as,
\beqa
c^{uu}_n = \frac{c^{n0}_n + c^{(n-2) 2}_n }{ 4}, \qquad{\rm and}\qquad
c^{ud}_n = \frac{c^{n0}_n - c^{(n-2) 2}_n }{4} .
\eeqa
The $2$-nd order flavour diagonal and off-diagonal susceptibilities are
given by,
\beqa
  \nonumber
  \frac{\chi_{uu}(T,\mu_u=0,\mu_d=0)}{T^2} &=& \l. \frac{\partial^2
  P(T,\mu_u,\mu_d)}{\partial\mu_u^2} \r|_{{\mu_u=\mu_d=0}} = 2c_2^{uu},
  \qquad{\rm and}\qquad \\ \nonumber
  \frac{\chi_{ud}(T,\mu_u=0,\mu_d=0)}{T^2} &=& \l. \frac{\partial^2
  P(T,\mu_u,\mu_d)}{\partial\mu_u\partial\mu_d} \r|_{{\mu_u=\mu_d=0}} = 2c_2^{ud} .
\eeqa

\begin{figure}[!tbh]
   {\includegraphics [scale=0.45] {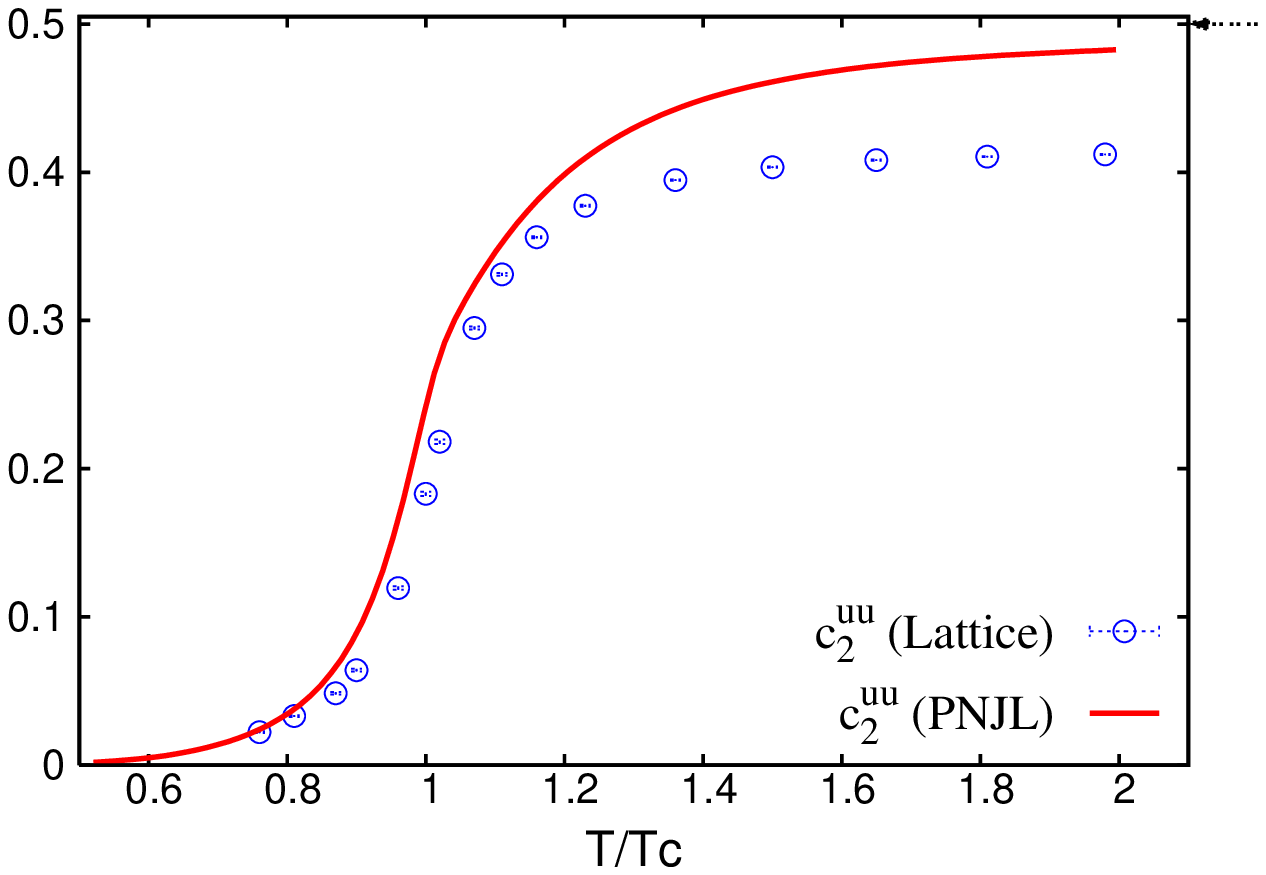}}
   {\includegraphics[scale=0.45]{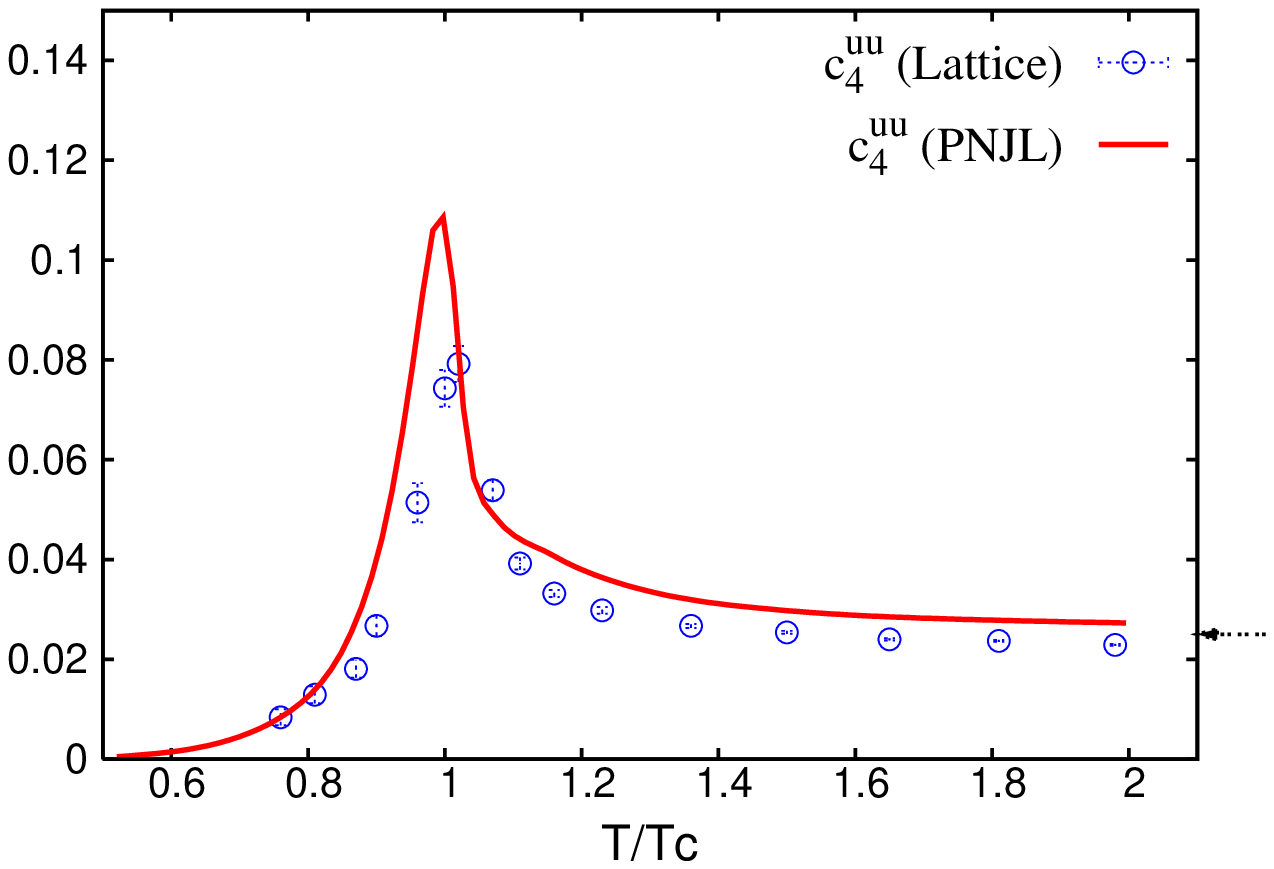}}
   {\includegraphics[scale=0.45]{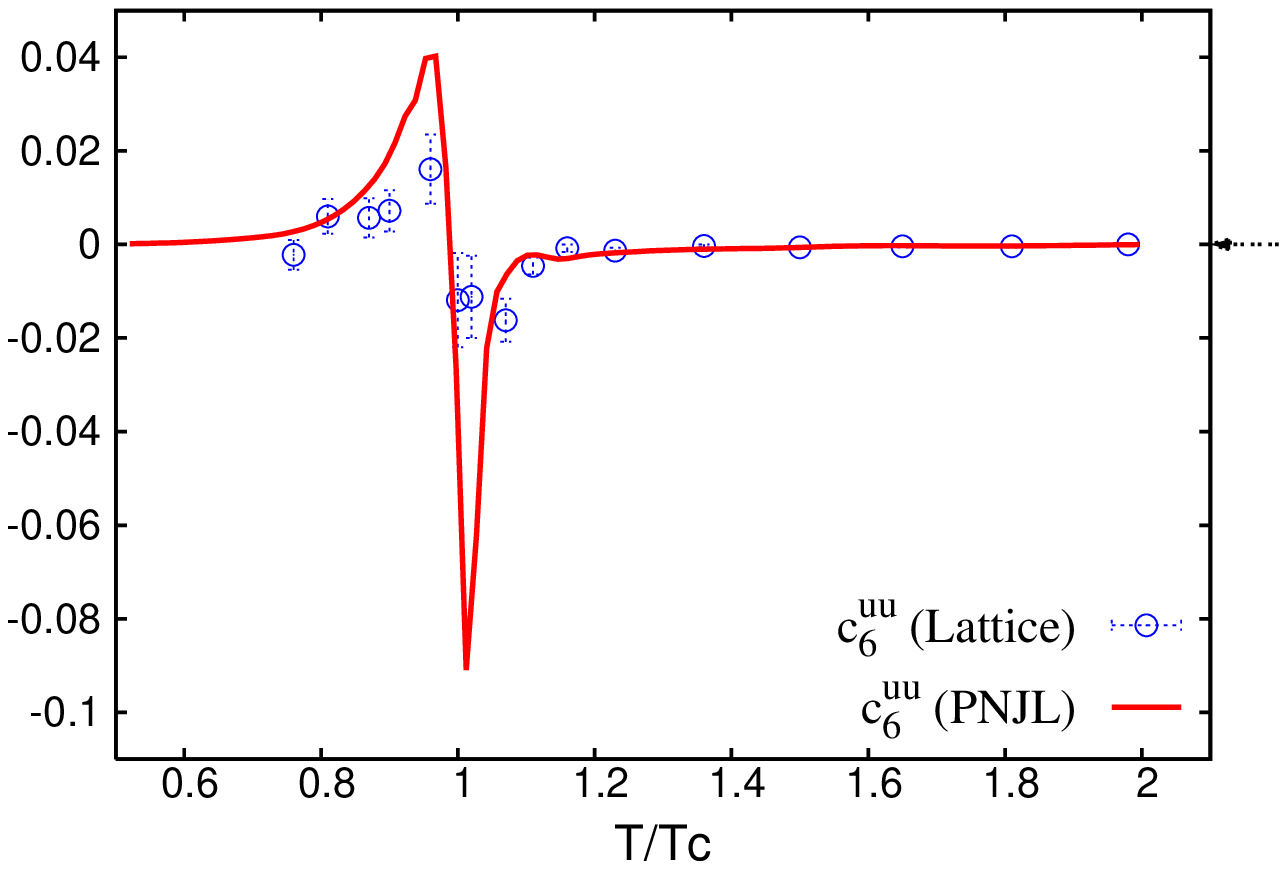}}
   {\includegraphics [scale=0.45] {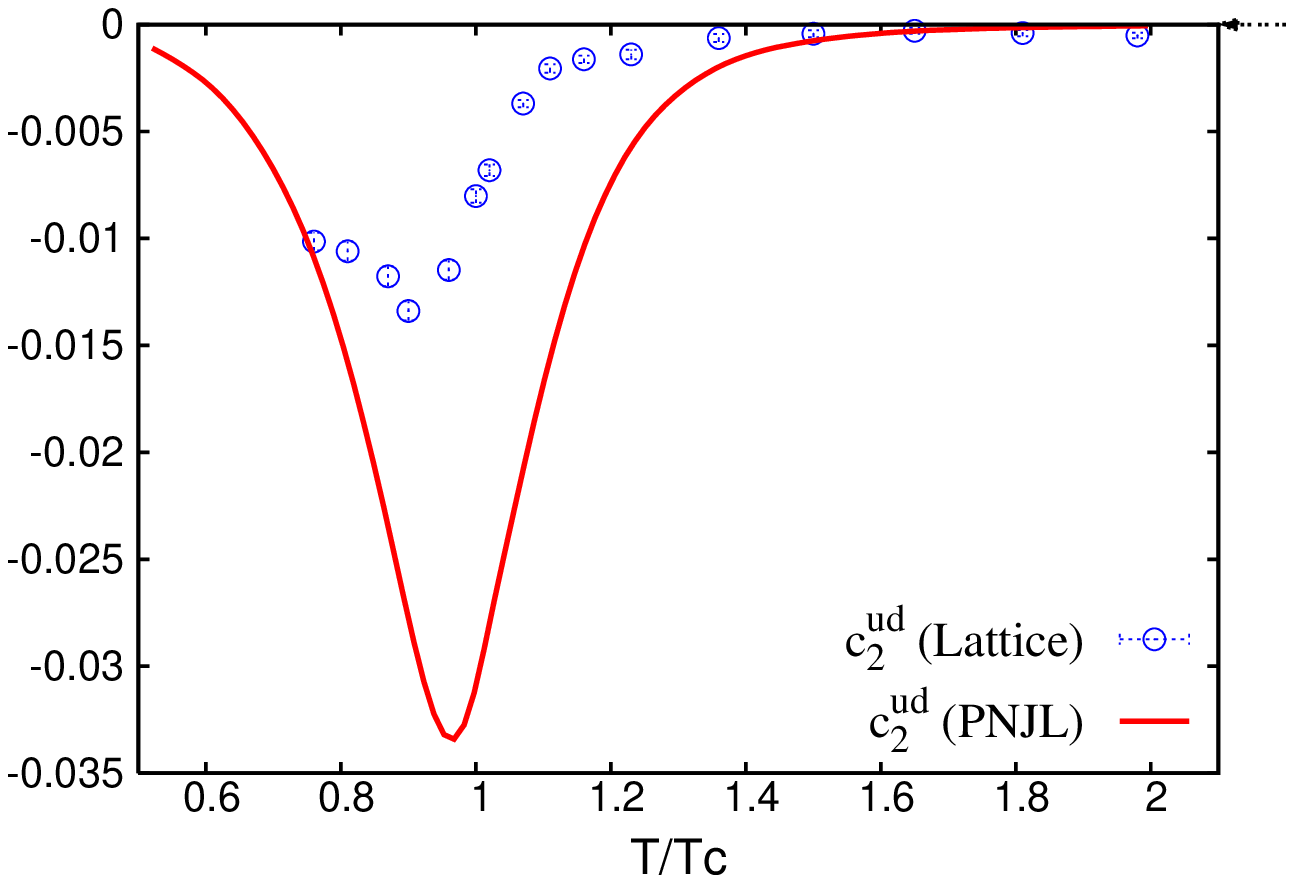}}
   {\includegraphics[scale=0.45]{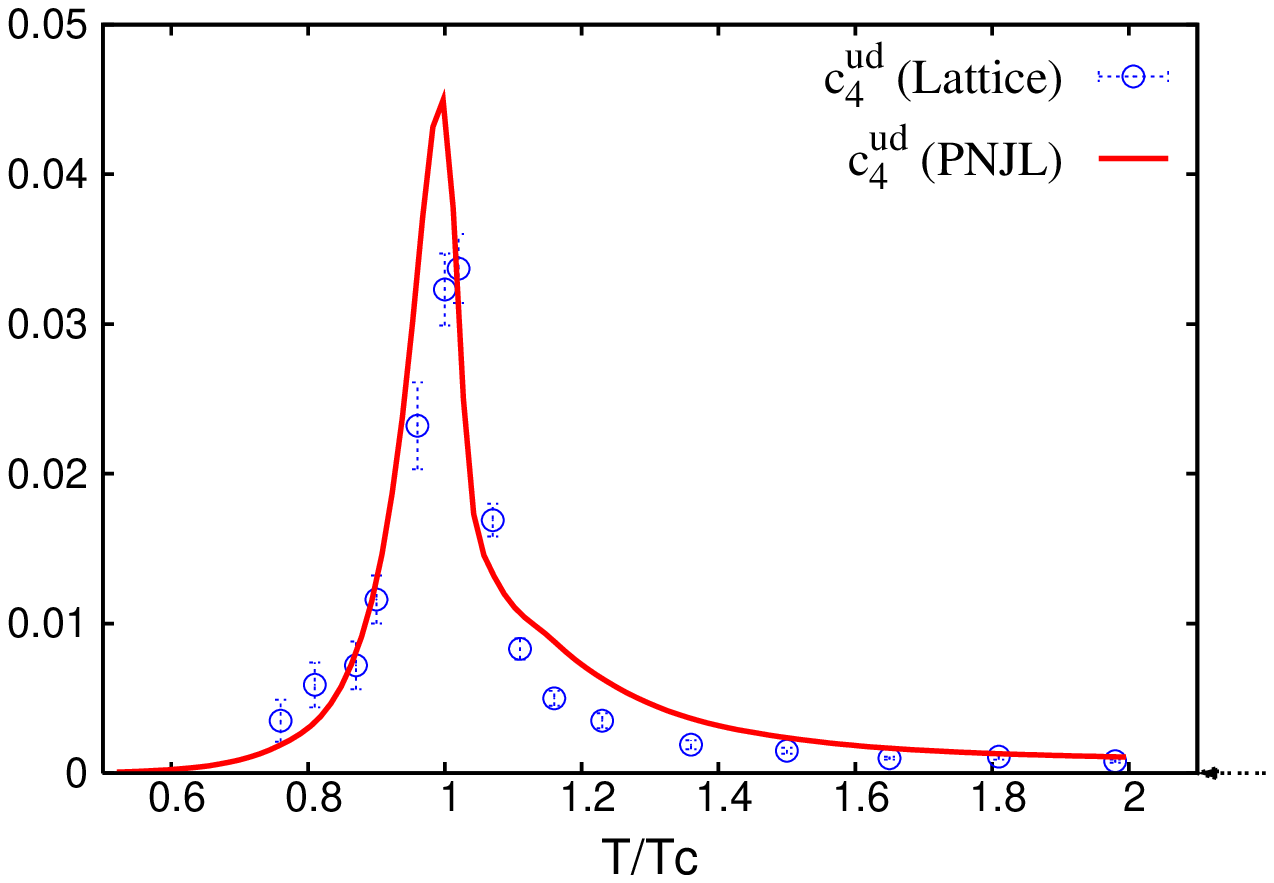}}
   {\includegraphics[scale=0.45]{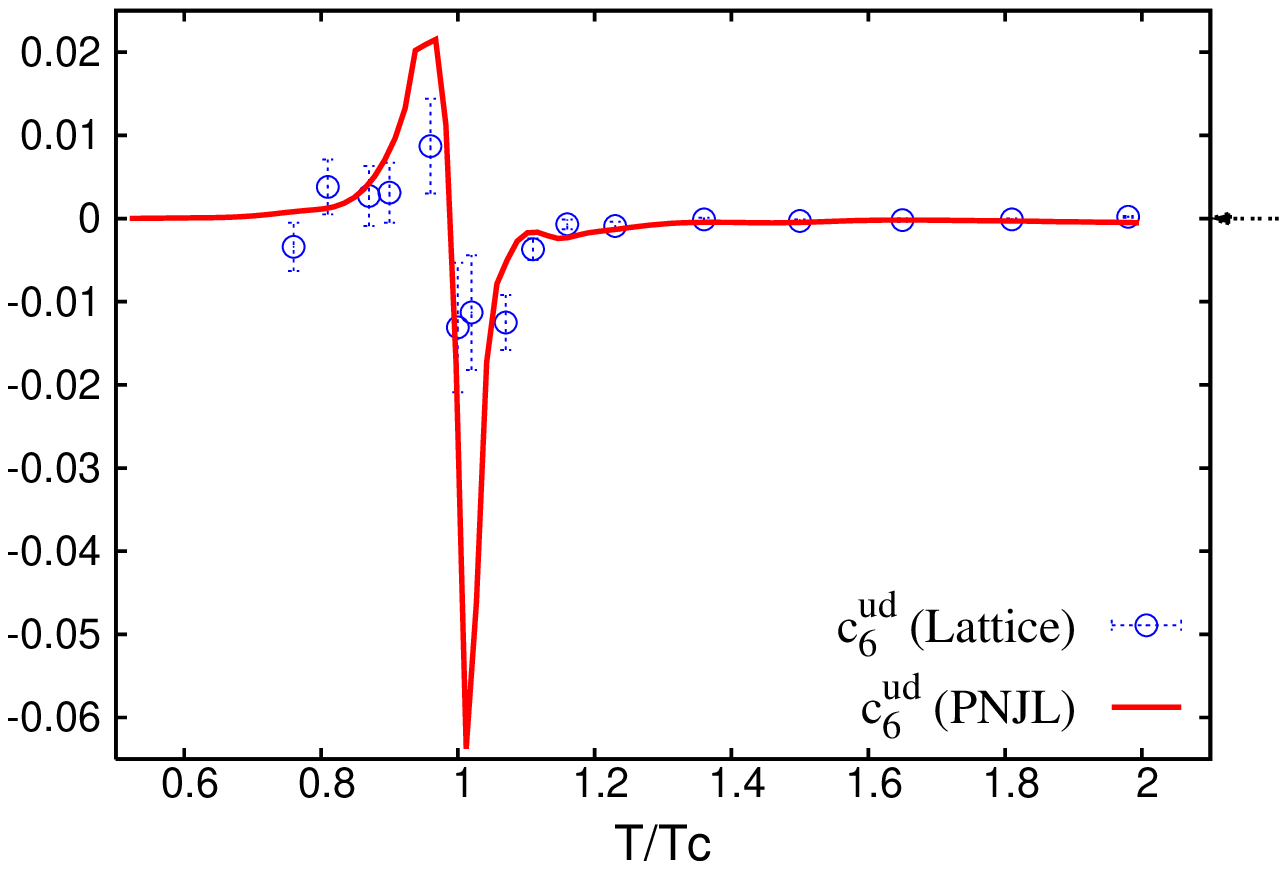}}
   \caption{The flavour diagonal ({\it upper row}) and flavour
   off-diagonal ({\it lower row}) susceptibilities for $n = 2$, $4$ and
   $6$ as functions of $T/T_c$. Symbols are LQCD data \cite{sixx}. The 
   arrows on the right indicate the respective ideal gas values.}
\label{fg.diodi}\end{figure}

These are shown in Fig.\ \ref{fg.diodi}.  Except $c_2^{uu}$,
all the other LQCD diagonal and off-diagonal coefficients are
close to their respective ideal gas values from $1.2T_c$ 
onwards. The most striking
discrepancy without the VdM term w.r.t the LQCD data was (see \cite{pnjl4}) 
in the $2$-nd order flavour off-diagonal susceptibility $c^{ud}_2$. 
$c_2^{ud}$ signifies the mixing of $u$ and $d$ quarks through the 
contribution of the two disconnected $u$ and $d$ quark loops. While the 
LQCD data shows that this kind of correlation between the $u$-$d$ flavours 
are almost zero just away form $T_c$, the PNJL model results remained
non-zero even upto $2 T_c$. Adding the VdM term this part of the PNJL
physics is now consistent with LQCD results. Below $1.2T_c$ there is
still a large quantitative difference between the PNJL and LQCD results
for $c_2^{ud}$. Obviously the VdM term is not expected to affect the
results at low temperatures significantly. At the moment it is not clear 
what physics lie
behind the difference between PNJL and LQCD results for $c_2^{uu}$ at
high temperatures and $c_2^{ud}$ at low temperatures. Perhaps the
quark masses may hold an answer. 

\section{Summary} \label{sc.summary}

In this work the PNJL model of Ref.\  \cite{pnjl2,pnjl3,pnjl4} has been 
extended by introducing a VdM term. The important change it brings 
about is to set the upper limit of the Polyakov loop to 1. With this 
model we have studied some thermodynamic properties of strongly 
interacting matter with the light flavours $u$ and $d$ within a 
certain range of temperature $T$, and small values of chemical 
potentials $\mu_0$ and $\mu_I$. In principle the VdM term affects
all thermodynamic quantities. We adjusted the parameters in the
model so that the pressure and energy density is close to that 
computed in LQCD. We have then made estimates of the specific heat,
the speed of sound and conformal measure. 

Further, we have extracted the Taylor expansion coefficients of 
pressure in the two chemical potentials upto $6$-th order. All 
the coefficients approach their respective SB limit above $2 T_c$.
A quantitative comparison with the LQCD results show reasonable
agreement, though the QNS $c_2$ and the INS $c^I_2$ on the Lattice 
are smaller by
about 20 $\%$. In contrast our earlier estimates \cite{pnjl3,pnjl4} 
of these coefficients without the VdM term showed that $c_4$ and
$c_I^2$ differ from the LQCD results. Thus the main effect of the 
VdM term is to impose physical constraints on $\Phi$ and 
$\bar{\Phi}$ such that at large temperatures the coefficients of
the same order approach each other. This is clearly visible from
the flavour off-diagonal coefficients shown in Fig. \ref{fg.diodi}.
The remaining difference of the values of the QNS and INS in the
model and lattice still needs to be addressed.
Possible future steps to bring in better agreement could be to
include beyond mean field effects and/or to include some sort of
temperature dependence to the coefficient of the VdM term.
However the lattice quark masses may be important in bridging
the gap. We already found that such data for pressure with almost
physical quark masses \cite{datta} show an increase at any given 
temperature when compared to data with larger quark masses \cite{leos}. 
This would encourage us to believe that extraction of the susceptibilities
with similar quark masses on the lattice may have a better agreement
with our results. Another way to compare results would be to
re-estimate the parameters of the NJL model directly from the
pion mass and decay constants from the lattice. We hope to 
undertake such studies in future.

In an alternative formulation of the PNJL model including the effect
of the VdM term, the coefficients $c_2$, $c_4$, $c_6$ and $c_8$ have
been calculated \cite{ratti4}. Surprisingly, we more or less agree 
with those results quantitatively. Apart from the fact that this
may be possible due to various adjustable parameters in both the
models, the main reason seems to be the small dependence of $\Phi$ 
and $\bar{\Phi}$ on the chemical potentials. The basic difference 
between the two approaches is in the use of the VdM potential.  
The VdM term is required to obtain the mean field solution of 
$\Phi$ and $\bar{\Phi}$. But as we have explained in the formalism
that it should not be included in the expression for pressure.
On the other hand in Ref. \cite{ratti4} apart from obtaining the 
mean fields the VdM term is included while calculating the value 
of pressure. The difference in the mean field treatment coupled
by almost same final results provide hints to the fact that mean field
treatment has certain shortcomings and is unable to settle issues
at hand. It would thus be worthwhile to look beyond.

\begin{acknowledgments}
We would like to thank A. Bhattacharya, S. Datta, S. Digal, S. Gupta, 
S. Mukherjee, P.B. Pal and R. Pisarski for many useful discussions 
and comments. We are thankful to A. Dumitru and O. Kaczmarek for useful 
discussions on the lattice computation of the Polyakov loop. 
\end{acknowledgments}

\appendix
\section{}
\label{ap.appn1}

\begin{figure}[!tbh]
\subfigure[]{
   {\includegraphics [scale=0.6] {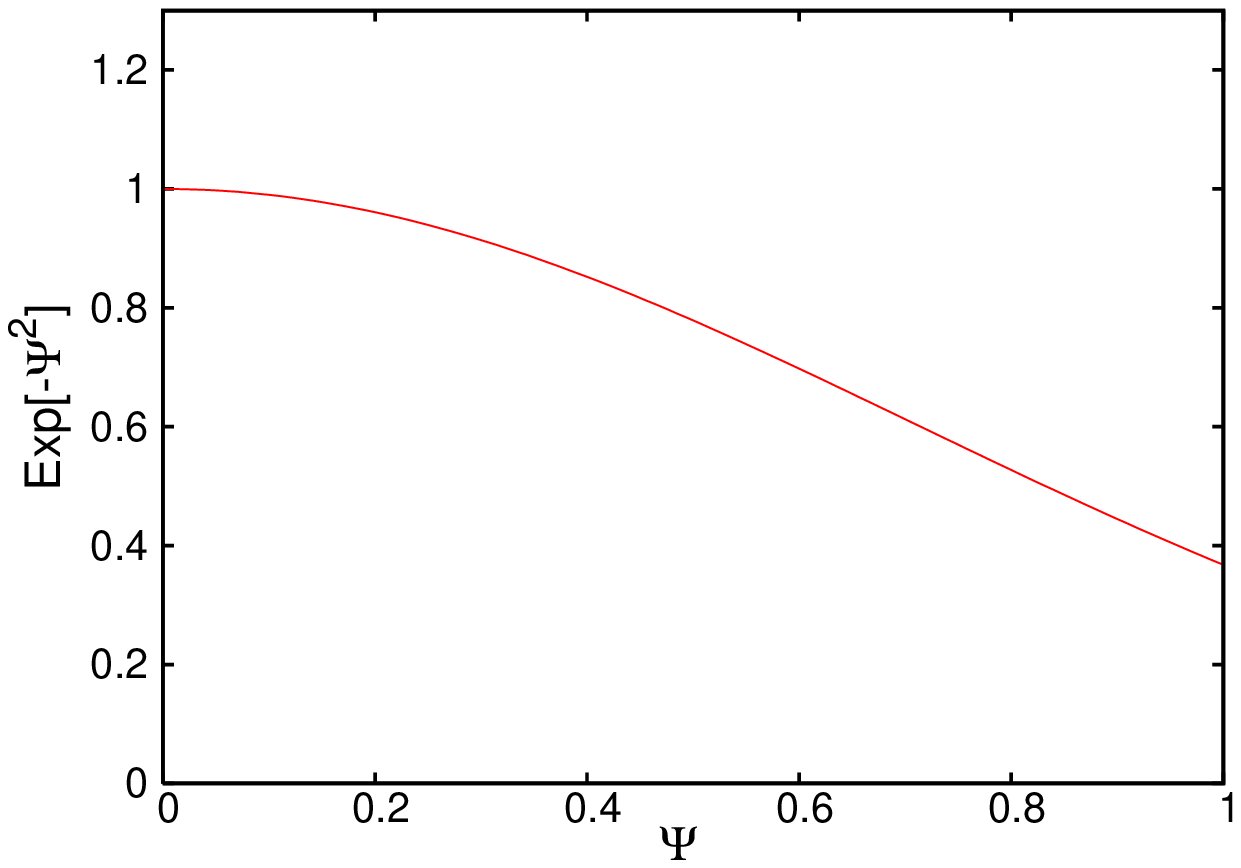}}
\label{fg.appn1}
}
\hskip 0.15 in
\subfigure[]{
   {\includegraphics[scale=0.6]{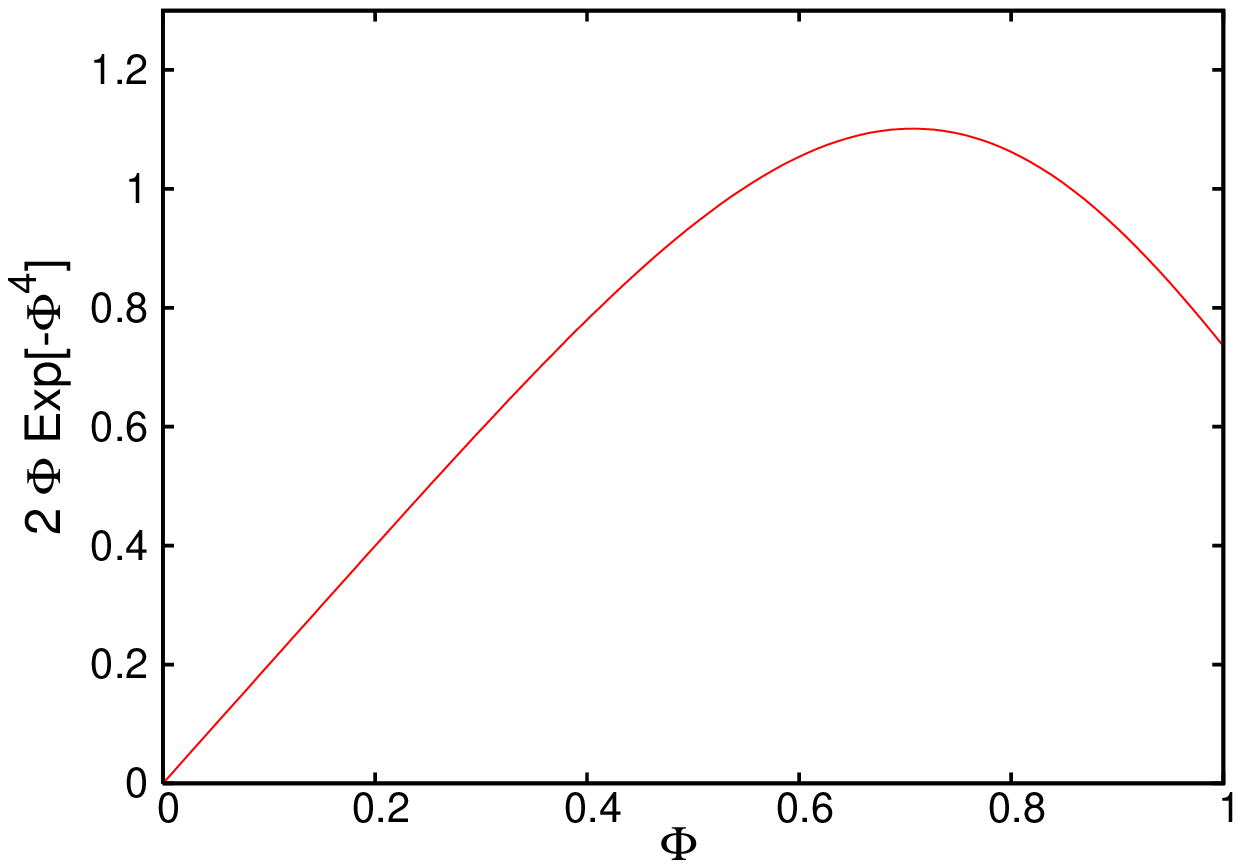}}
\label{fg.appn2}
}
   \caption{(a) $\Phi$ and (b) pressure for $\kappa = 0$($T_0 = 0.27$ GeV)
 and $\kappa=0.5$($T_0=0.2555$ GeV). The value of $T_c$ is 0.270 GeV.
   }
\end{figure}

We demonstrate the methodology of extracting the average value
of a quantity from mean field approximation. Suppose we have
a variable $\Psi$ with a probability distribution given by
$\exp(-\Psi^2)$ and we have to obtain the average of the function
in the exponential $\Psi^2$ (just like we have to obtain the
average of $\cV$ in Eqn. \ref{eq.prest}). We define the distribution 
in the domain $0\le \Psi \le 1$. The distribution is shown in 
Fig.\ref{fg.appn1}. The average is obtained as,

\beqa
\langle \Psi^2 \rangle = \frac{\int_0^1 d\Psi \Psi^2 e^{-\Psi^2}}
            {\int_0^1 d\Psi e^{-\Psi^2}} = 0.2537
\label{eq.appn1}
\eeqa

\noindent
where $Z = \int_0^1 d\Psi e^{-\Psi^2} = 0.747$ is like a partition
function. In the given domain the distribution has no maximum and 
thus a mean field solution cannot be obtained. 
Let us now make a change of variable from $\Psi$ to $\Phi$ where
$\Psi = \Phi^2$. The distribution becomes $2\,\Phi\,\exp(-\Phi^4)$ as
shown in Fig.\ref{fg.appn2}. Here, $2 \Phi$ is like the Jacobian
in the main text. 
One can now easily check that corresponding to Eqn.\ref{eq.appn1},
we need to find the expectation value of $\Phi^4$ given by,

\beqa
\langle \Phi^4 \rangle = \frac{\int_0^1 2 \,\Phi \,d\Phi \,\Phi^4\, e^{-\Phi^4}}             {\int_0^1 2\,\Phi\,d\Phi\,e^{-\Phi^4}} = 0.2537
\label{eq.appn2}
\eeqa

\noindent
In this case the distribution has a maximum and we can 
do a saddle point approximation. We thus minimize
$\Phi^4 - \ln [2 \Phi]$ which gives the mean field value
$\langle \Phi \rangle = 1/\sqrt{2}$. Using this value 
we find $\langle \Phi \rangle^4 = 0.25 \simeq \langle \Phi^4 \rangle$.
So $\langle \Phi \rangle^4$ gives a good approximation to 
$\langle \Phi^4 \rangle$. On the other hand if we include
the logarithm term we have,
$\langle \Phi \rangle^4 - \ln [2 \langle \Phi \rangle]
= -0.0966$, which is widely different. 


\end{document}